%% file: main.tex
\documentclass[letterpaper,twocolumn,10pt]{article}
\usepackage{usenix-2020-09}

\input{setup/setup}

\begin{document}
\title{\System{}: \SystemFull{} for Inter-Domain Networks}

\input{setup/authors} \maketitle

\blfootnote{This is the author’s version of the work, which, compared to the definitive version of record published at USENIX Security '23, includes an additional user study and an extended related work section.}

\input{sections/abstract}
\input{sections/introduction}
\input{sections/background}
\input{sections/model}
\input{sections/system}
\input{sections/specification}
\input{sections/implementation}
\input{sections/evaluation}

\input{sections/analysis}
\input{sections/discussion}
\input{sections/relatedwork}
\input{sections/conclusion}
\input{sections/acknowledgment}

\bibliographystyle{plain}
\typeout{}
\bibliography{bib/ref}
\appendix
\input{sections/appendix}

\end{document}

%% file: setup/setup.tex
\usepackage{amsmath,amssymb,amsfonts,amsthm}
\usepackage{balance}
\usepackage[noabbrev,capitalize]{cleveref}
\usepackage{enumitem}
\usepackage{pgfplots}
\usepackage[breakable]{tcolorbox}
\usepackage[binary-units,per-mode=symbol]{siunitx}
\usepackage[mode=buildnew,subpreambles]{standalone}
\usepackage{booktabs}
\hypersetup{colorlinks=true,urlcolor=cyan}
\usepackage{subcaption}
\usepackage{makecell}

\newcommand{\key}[2]{\ensuremath{\textsc{K}_{#1}^{#2}}}
\newcommand{\mac}[2]{\ensuremath{\textsc{MAC}_{#1}(#2)}}
\newcommand{\hvf}[1]{\ensuremath{\textsc{V}_{#1}}}
\newcommand{\aes}[2]{\ensuremath{\textsc{AES}_{#1}(#2)}}
\newcommand{\prf}[2]{\ensuremath{\textsc{PRF}_{#1}(#2)}}
\newcommand{\tspkt}[0]{\ensuremath{\mathrm{tsPkt}}}
\newcommand{\ind}[2]{\ensuremath{\psi_{#1}^{#2}}}
\newcommand{\lval}[0]{\ensuremath{\ell_\text{val}}}
\newcommand{\lpol}[0]{\ensuremath{\ell_\text{pol}}}

\newcommand{\system}{FABRID}
\newcommand{\System}{FABRID}
\newcommand{\SystemFull}{Flexible Attestation-Based Routing}

\newcommand{\enumfont}{\bfseries}
\SetEnumitemKey{endpoint}{label={\enumfont E\arabic*}, ref=E\arabic*}
\SetEnumitemKey{as}{label={\enumfont A\arabic*}, ref=A\arabic*}

\definecolor{blue}{HTML}{007A92}
\definecolor{purple}{HTML}{91056A}
\definecolor{green}{HTML}{72791C}
\definecolor{red}{HTML}{A8322D}

\pgfplotsset{compat=1.17}
\pgfplotscreateplotcyclelist{eval list fabridcyclelist}{
	blue, every mark/.append style={solid,fill=blue},mark=halfsquare*\\
	purple, every mark/.append style={solid,fill=purple},mark=triangle*\\
	green, every mark/.append style={solid,fill=green},mark=*\\
	red, every mark/.append style={solid,fill=red},mark=square*\\
	blue, dashed, every mark/.append style={solid,fill=blue},mark=halfsquare*\\
	purple, dashed, every mark/.append style={solid,fill=purple},mark=triangle*\\
	green, dashed, every mark/.append style={solid,fill=green},mark=*\\
	red, dashed, every mark/.append style={solid,fill=red},mark=square*\\
}

\tcbuselibrary{breakable}
\newcommand{\greyborder}[1]{%
    \begin{tcolorbox}[breakable, colback=white, colframe=darkgray, left=1pt,right=10pt,top=0pt,bottom=0pt]%
    #1%
    \end{tcolorbox}%
}

\newcommand\blfootnote[1]{%
  \begingroup
  \renewcommand\thefootnote{}\footnote{#1}%
  \addtocounter{footnote}{-1}%
  \endgroup
}
\renewcommand\footnoterule{\kern-3pt \hrule width \linewidth \kern 2.6pt}

%% file: setup/authors.tex
\author{
{\rm Cyrill Krähenbühl}\\
\hspace{1cm}ETH Zürich\hspace{1cm} \and
{\rm Marc Wyss}\\
\hspace{1cm}ETH Zürich\hspace{1cm} \and
{\rm David Basin}\\
\hspace{1cm}ETH Zürich\hspace{1cm} \and
{\rm Vincent Lenders}\\
\hspace{1cm}armasuisse\hspace{1cm} \and
{\rm Adrian Perrig}\\
\hspace{1cm}ETH Zürich\hspace{1cm} \and
{\rm Martin Strohmeier}\\
\hspace{1cm}armasuisse\hspace{1cm}
}

%% file: sections/abstract.tex
\begin{abstract}
In its current state, the Internet does not provide end users with transparency and control regarding on-path forwarding devices.
In particular, the lack of network device information reduces the trustworthiness of the forwarding path and prevents end-user applications requiring specific router capabilities from reaching their full potential.
Moreover, the inability to influence the traffic's forwarding path results in applications communicating over undesired routes, while alternative paths with more desirable properties remain unusable.

In this work, we present \system{}, a system that enables applications to forward traffic flexibly, potentially on multiple paths selected to comply with user-defined preferences, where information about forwarding devices is exposed and transparently attested by autonomous systems (ASes).
The granularity of this information is chosen by each AS individually, protecting them from leaking sensitive network details, while the secrecy and authenticity of preferences embedded within the users' packets are protected through efficient cryptographic operations.
We show the viability of \system{} by deploying it on a global SCION network test bed, and we demonstrate high throughput on commodity hardware.
\end{abstract}

%% file: sections/introduction.tex
\section{Introduction}
The current Internet can be viewed as a black box over which endpoints send traffic without detailed knowledge about the network paths taken.
While this black box usage is sufficient for many applications, it can lead to performance, compliance, and security issues when the path does not comply with end-users'  required network properties.

For example, some applications require that data does not leave a given jurisdiction, which therefore requires that sensitive traffic is only routed through routers operated in certain countries or regions~\cite{EuropeanParliament2016}.
Another application is time synchronization, which may require packets to be forwarded only over routers with hardware-based time synchronization support (e.g., PTP~\cite{PTP2019})~\cite{Salazar2019}.
Governments or critical infrastructure operators may further desire to avoid network equipment from certain manufacturers or specific router versions that pose potential security risks.
They may thus wish to route traffic over paths consisting of specific, trusted equipment.

Confidentiality of data in the Internet is achieved through encryption.
However, data (payload) encryption has limitations.
Metadata such as IP addresses can still be observed, enabling censorship and reducing anonymity.
The distribution and protection of keys is challenging and weaknesses in ciphers and modes of encryption, advances in cryptanalysis, or quantum computers may render encryption schemes insecure.
Moreover, legacy communication protocols might not support encryption at all.
Instead of exclusively relying on encryption, confidentiality and integrity can be enhanced by forwarding data over channels built from trusted network equipment (devices and links), which is believed not to eavesdrop on the communication.

In a survey we conducted last year, we learned that there is a desire by network and security experts to be able to select paths in the Internet based on particular router characteristics. Namely, over \SI{80}{\percent} of the surveyed end-users currently miss fine-grained path transparency in the Internet.
Over \SI{60}{\percent} of end-users would be even willing to pay more to their Internet Service Provider (ISP) in order to be able to control path characteristics such as network operator, jurisdiction, geolocation of network operators and routers, runtime status, and router hardware manufacturer.
This raises the question of how to implement inter-domain Internet path selection for end-users based on attested router properties.

There are currently two initiatives with similar design goals, however neither of them can provide flexible, policy-compliant, inter-domain Internet path selection based on individual user needs.
First, the IETF draft Trusted Path Routing (TPR)~\cite{voit-rats-trustworthy-path-routing-06} from the Remote ATtestation ProcedureS (RATS) working group~\cite{rats} has proposed to conduct remote attestation measurements between neighboring routers to find trusted links, and thus create a trusted routing plane within a network based on intra-domain routing protocols such as IS-IS or OSPF\@.
Unfortunately, TPR does not offer endpoint path control and focuses on intra-domain communication.

The second approach is the path-aware Internet architecture SCION~\cite{Chuat2022a}, which allows individual applications, on endpoints, to influence the forwarding path of their traffic.
SCION is an Internet architecture that enables endpoint path control and multi-path communication.
SCION is now globally deployed, commercially available, and allows endpoints to control forwarding at the autonomous system (AS) level.
Still, the level of transparency and control provided by SCION does not suffice for an endpoint to learn relevant information about network equipment in the internal networks of on-path ASes, as it does not provide attested router properties for the network elements along a selected path.

In this work, we present \SystemFull{} (\system{}), a novel system that supports the fine-grained selection of the properties of all network elements along an inter-domain path.
ASes advertise network devices with specific properties and optionally increase their trustworthiness by publishing remote attestation results to prove the existence of the advertised devices.

\System{} extends SCION, inheriting its properties such as path control at the granularity of border routers, robustness against single link failures, and a scalable global route distribution mechanism.
We leverage the backward-compatible extensibility of SCION's control and data planes to enable partial deployment of \system{} within the SCION network.
Even when some ASes on the forwarding path do not support it, \system{} still provides some benefits.
Experience with IPv6~\cite{ipv6_statistics}, egress filtering, and BGPsec~\cite{ripe_statistics} shows that allowing incremental deployment is a necessity for new Internet technologies since convincing ISPs to adopt them is a challenging and time-consuming process.

The main contributions of this work are the following:
\begin{itemize}
    \setlength\itemsep{0pt}
    \item We perform a user study to evaluate the interest of users in various router attributes and whether network operators would be willing to reveal such attributes.
    \item We design an extensible policy language to describe arbitrary router attributes, which supports customizable policies for users and ASes.
    Leaking sensitive intra-AS data is prevented by restricting AS policies accordingly.
    \item We enhance SCION's path control to enable path selection based on user-defined policies and use SCION's public-key infrastructure to authenticate information about on-path network devices by the respective ASes.
    \item We apply efficient measures for authenticity and secrecy of users' packet-carried policy decisions.
    \item We implement and evaluate a high-speed variant of \system{}, and demonstrate the system's feasibility by deploying it on a global test bed.
\end{itemize}

%% file: sections/background.tex
\section{Background on SCION}\label{sec:background}

\begin{figure}[t]
  \includegraphics[width=\linewidth]{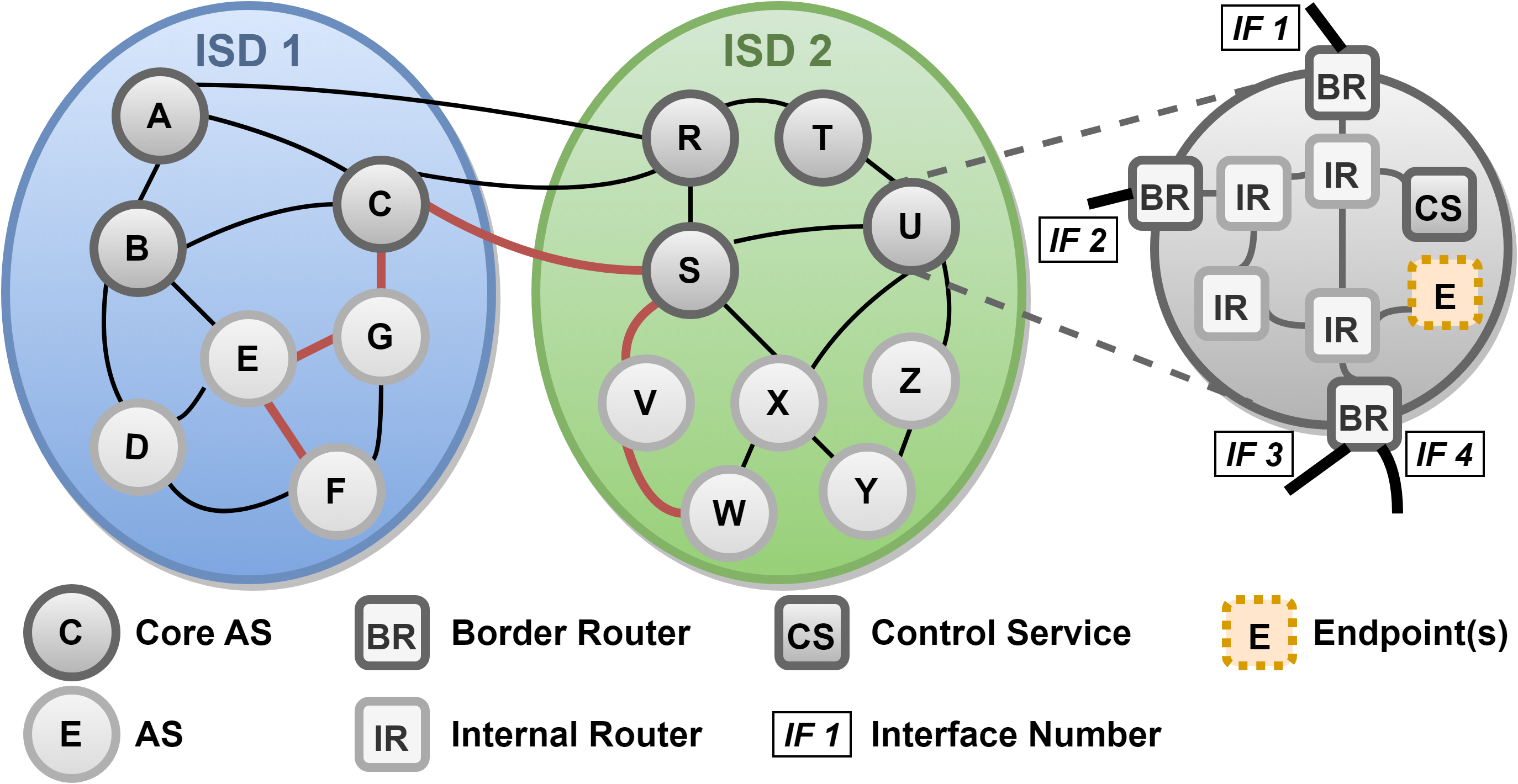}
  \caption{SCION topology example.
    Assuming the existence of three path segments C-G-E-F, C-S, and S-V-W, these three path segments can be combined into an end-to-end path from a host in AS F to a destination host in AS W.
  }\label{fig:scion_overview}
\end{figure}

We build \system{} on top of SCION~\cite{Chuat2022a}, a network architecture that provides strong availability, control, scalability, and transparency guarantees.
In SCION, autonomous systems (ASes), which constitute the Internet's building blocks, run a dedicated SCION control service responsible for the discovery, dissemination, and validation of routing information.
SCION border routers provide one or multiple \emph{external interfaces} to other ASes, connecting the network link between them, and an \emph{internal interface} for connectivity to border routers in the same AS, the control service, and other local endpoints.
All other routers inside an AS are referred to as internal routers.

In SCION, ASes are logically grouped into isolation domains (ISDs).
Some ASes inside an ISD are core ASes, which manage the ISD's trust roots and provide connections to core ASes in the same or other ISDs.
Routing in SCION is a per-ISD process for discovering intra-ISD paths.
To interconnect ASes in different ISDs, a separate routing process discovers paths between all core ASes.
Intra- and inter-ISD routing procedures are implemented through a beaconing process, which is initialized by core ASes disseminating path-segment construction beacons (PCBs).
An AS receiving a PCB extends it with local information such as its AS and interface identifiers.
An AS signs the PCB and then forwards in to selected neighbors.
The AS's public key is certified by SCION's control-plane public-key infrastructure (\mbox{CP-PKI}).
SCION enables a diversity of end-to-end paths thanks to the various ways in which up to three path segments can be combined: an up-segment from its AS to a core AS of the same ISD, a core-segment between two core-ASes, and a down-segment between a core- and the destination AS.
Shortcuts between up- and down-segments eliminate potential routing inefficiencies.

An endpoint learns segments to the desired destination from the control service of its AS, where the service typically returns a few dozen possible segments.
This makes SCION a path-aware networking architecture, as endpoints can influence the forwarding path that their traffic takes through the network on an AS-level.
This is in contrast to traditional inter-domain networking, where the network  delivers data on paths outside of the endpoints' control.

When sending a packet, the path (a list of ingress/egress interface-pairs) is encoded as packet-carried forwarding state (PCFS) in the packet header, instructing on-path border routers how to forward the packet.
PCFS enforces the endpoints' path selection, enables multipath communication, and protects packets from unanticipated re-routing and hijacking attacks.
An example SCION topology is shown in \cref{fig:scion_overview}.

%% file: sections/model.tex
\section{Trust Model and System Objectives} \label{sec:model}
In this work, we consider \emph{endpoints} that require communication with other endpoints over paths consisting of devices with specific desired attributes.
We intentionally keep the definition of an endpoint abstract---it can refer to a machine, an application, or even a specific flow.

\subsection{System Objectives} \label{sec:model:system_objectives}
\newtheorem{property}{Property}
\newtheorem*{idealobjective}{Ideal Objective}
\newtheorem*{realisticobjective}{Realistic Objective}

\system{}'s overall objective is to enable endpoints to send traffic over devices or links with desired attributes.
We want to support a large variety of use cases, from utilizing specific functionality provided only by certain routers to ensuring that traffic is sent only over trusted hardware.
The generalized ideal property of \system{} is the following:
\greyborder{
\begin{idealobjective}\label{property:optimal}
  Traffic is only received by devices or links with attributes acceptable to the endpoints.
\end{idealobjective}
}
In practice, achieving this property is challenging for multiple reasons.
Network operators would need to give external entities, e.g., endpoints, insight into communication devices and channels on \emph{all layers}, i.e., on the network-, link-, and physical layer, which might not readily be available.
Even given this insight, adjusting the routing of the typically static link-, and physical layer can be difficult.
At the same time, routing at the network layer offers more flexibility and is very common, e.g., in traffic engineering, especially given the prevalence of software-defined networking (SDN).
Additionally, control over the link-, and physical layer path has limited benefits as the link-, and physical layer path often depends on the network path.
Finally, for the trusted hardware use case, all of these devices would require hardware support, e.g., a trusted platform module (TPM).
We therefore relax this ideal property as follows:
\greyborder{
\begin{realisticobjective}\label{property:reduced}
  Traffic is only received by network layer devices (routers) with attributes acceptable to the endpoints.
\end{realisticobjective}
}
This is still a very strong property, as the physical- and link layers are not easily accessible to an adversary.
Link layer devices cannot be addressed directly by arbitrary devices on the Internet, and eavesdropping on links is difficult due to physical security measures such as protected premises at ISPs and Internet Exchange Points (IXPs).

\newcommand{\questionnairefigheight}{5cm}
\begin{figure*}[t]
  \captionsetup[subfigure]{margin={1.8cm,0.1cm}}
  \begin{subfigure}[t]{.54\linewidth}
    \includegraphics[height=\questionnairefigheight{}]{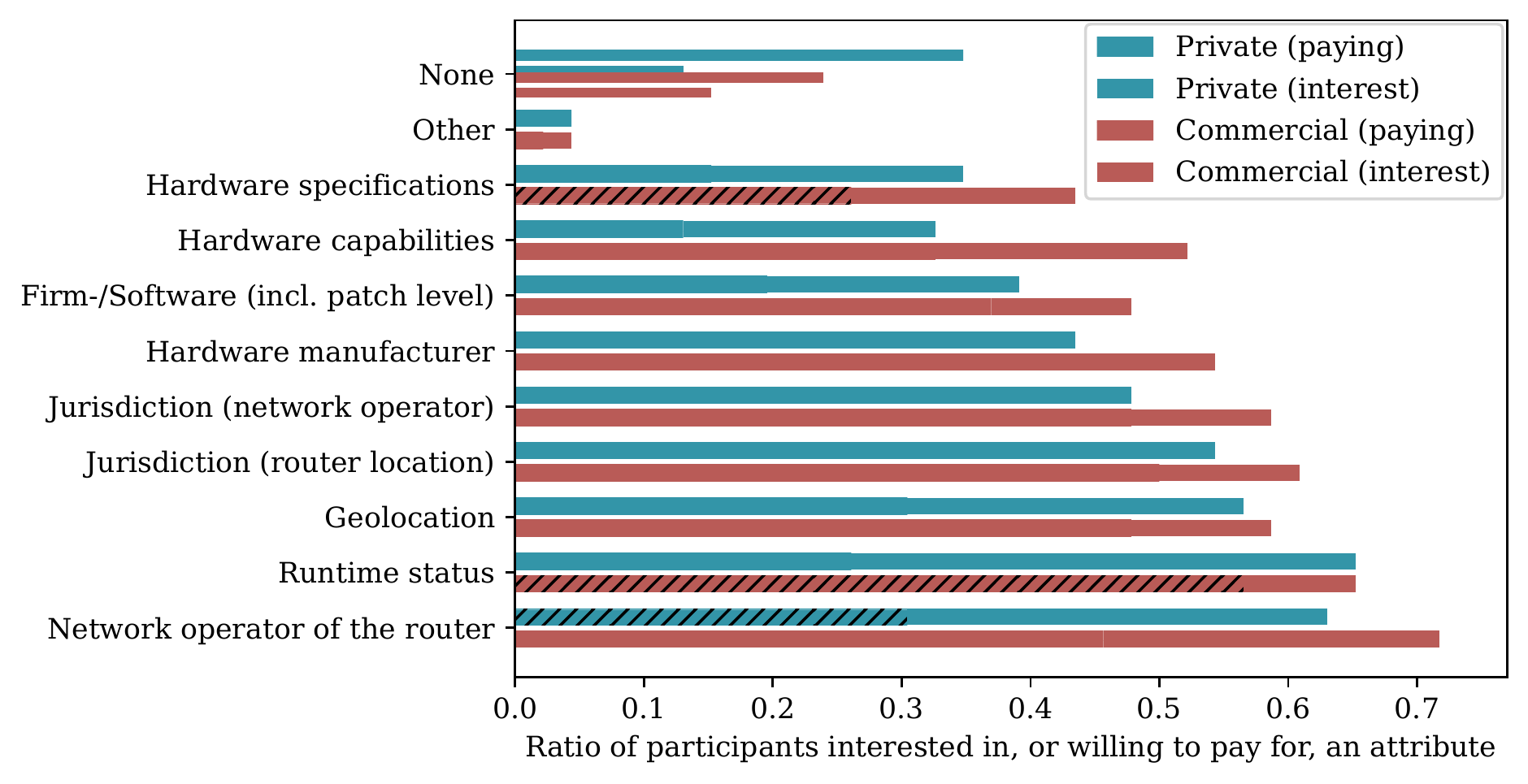}
    \caption{Desirable router attributes from the perspective of a private and commercial customer and their willingness to pay for only sending traffic over desirable routers (all 46 participants).}\label{fig:questionnaire:user-answers}
  \end{subfigure}
  \hspace{1mm}
  \captionsetup[subfigure]{margin={0.1cm,0cm}}
  \begin{subfigure}[t]{.39\linewidth}
    \includegraphics[height=\questionnairefigheight{}]{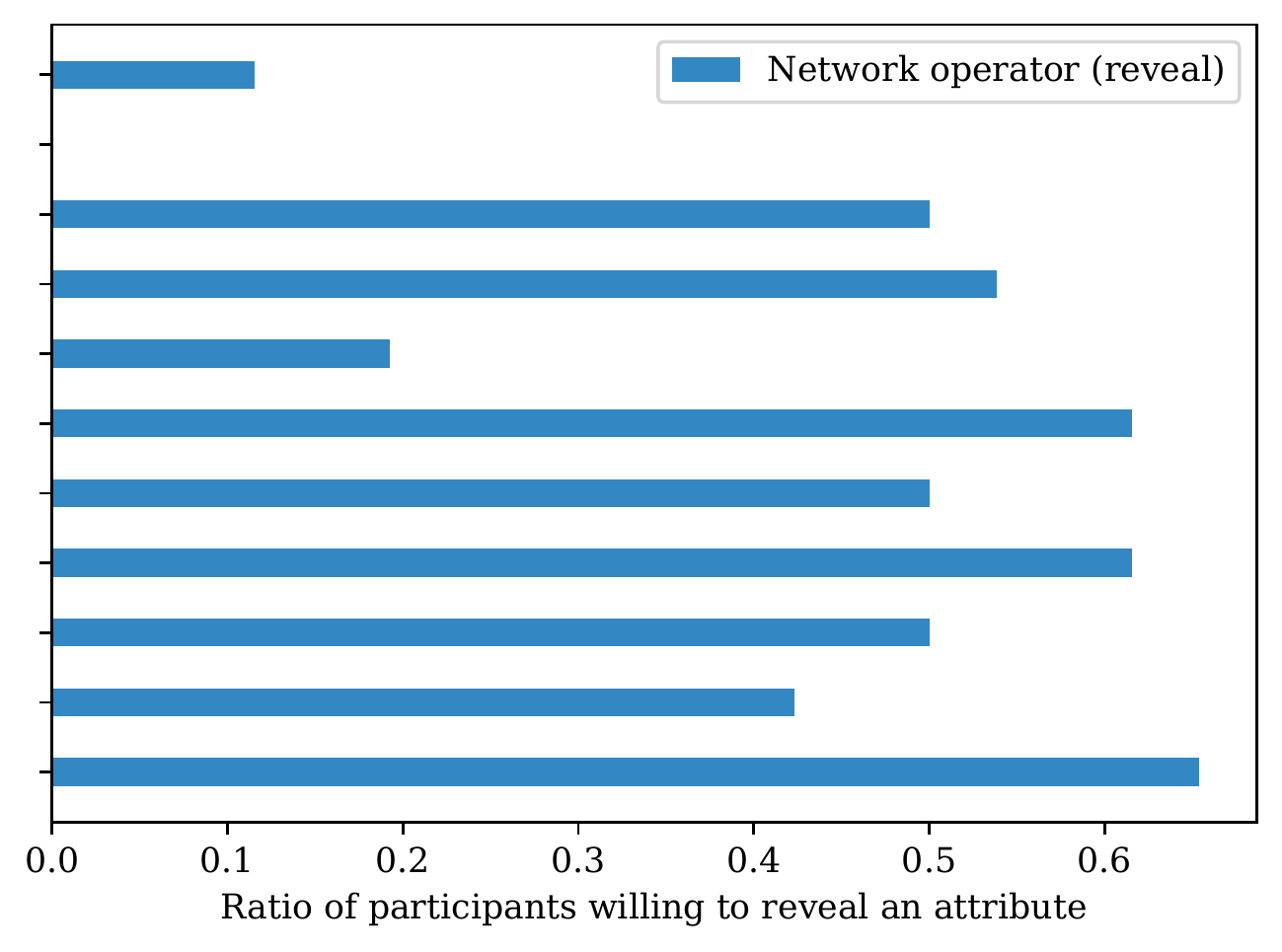}
    \caption{Router attributes a network operator is willing to reveal to customers (6 participants with ISP background).}\label{fig:questionnaire:netop-answers}
  \end{subfigure}
  \caption{The relevance of router attributes from different points of view based on the questionnaire results.}\label{fig:questionnaire}
\end{figure*}

To achieve this objective we formulate the following endpoint requirements:
\begin{enumerate}[endpoint]
    \setlength\itemsep{0pt}
    \item \label{endpoint_transparency} \textbf{Path Transparency}:
      Endpoints can request information on the attributes of on-path devices. Accurate information allows them to select paths to achieve specific objectives.
    \item \label{endpoint_attestation} \textbf{Attested Claims}:
    Claims about router attributes should be attested by the AS (and optionally by the routers themselves) to the endpoints and be non-repudiable.
    \item \label{endpoint_privacy} \textbf{Secrecy and Authenticity}:
    For packets that are meant to follow a path with certain attributes, only legitimate on-path entities should be able to decrypt those attributes embedded in the packet header.
    Those entities should also verify the authenticity of the attributes and discard packets not passing this verification.
    \item \label{endpoint_path_stability} \textbf{Path Stability}:
    A selected path consisting of routers with desired attributes should be stable over time, otherwise traffic might be routed over devices violating the endhost's preference.
    \item \label{endpoint_path_validation} \textbf{Path Validation}:
    Even if path stability is ensured by some routing protocol, misconfigured or compromised devices may still violate their forwarding directives.
    It should therefore be possible for endpoints to verify that their selected path is actually obeyed by on-path routers.
    \item \label{endpoint_fine_grained} \textbf{Fine-Grained Properties}: 
    Specifying desired router attributes should be as flexible as possible, irrespective of endpoints being end hosts, applications, or flows.
\end{enumerate}
Note that some of these requirements are not possible to achieve in the current Internet based on the Border Gateway Protocol (BGP).
The most fundamental limitation here is BGP's lack of path stability, whereby routes can unpredictably change at the level of ASes due to rerouting events or hijacking attacks.
ASes have fundamentally different requirements:
\begin{enumerate}[as]
    \setlength\itemsep{0pt}
    \item \label{ases_secrets} \textbf{Trade Secrets}: 
    An AS should not need to reveal sensitive information about its internal network, but rather be able to decide itself which router attributes to publish.
    \item \label{ases_ddos_protection} \textbf{DDoS Protection}:
    Some ASes distribute ingress traffic for the same egress router among multiple paths to achieve better redundancy and higher throughput.
    Such ASes should have the option to announce router attributes without having to specify a concrete path, as otherwise an adversary can target this path using DDoS.
    \item \label{ases_efficient_distribution} \textbf{Efficient Distribution}: 
    Propagating router attributes to endpoints should be efficient, i.e., incur only low computation and communication overhead.
    \item \label{ases_efficient_forwarding} \textbf{Efficient Forwarding}:
    Sending traffic along routers with specific attributes should not significantly impact performance. In particular, this implies efficient packet generation and reception at the endpoints and high-speed forwarding at routers that scales to today's Internet link capacities. Also, the amount of additional packet meta data should be minimal.
\end{enumerate}
We address the tension between the transparency requirements of endpoints (\ref{endpoint_transparency}) and what router information an AS is willing to provide (\ref{ases_secrets}) in two ways.
First, we conduct a survey to discover what router attributes are desired by endpoints and what attributes AS operators are willing to make public.
Second, we introduce the concept of policies, which allows ASes to precisely specify the granularity of attributes to be published.

\subsection{Survey}\label{sec:survey}
We conducted a survey to analyze what router attributes private and commercial customers are interested in, whether they are willing to pay for sending traffic only over routers with desirable attributes,
and which router attributes network operators are actually willing to reveal.
In total, our survey was answered by \num{46} participants which are knowledgeable in routing and security.
\Cref{fig:questionnaire} shows the survey's main results.
First, we observe a large overlap in router attributes that are desired by the participants and attributes that network operators are willing to reveal.
Examples are the router's network operator and its jurisdiction.
However, there is also a clear discrepancy for some attributes.
Roughly \SI{40}{\percent} of private and commercial customers are interested in routers' firm- and software, but most network operators are reluctant to share this information.
In contrast, network operators are willing to disclose information about the hardware used in their network, but this information is only relevant for some customers.
See \cref{sec:appendix:questionnaire} for a detailed discussion of the survey.

\subsection{Policies}
In this section, we introduce the concept of \emph{policies}. Policies allow ASes to specify the granularity of the router information they publish (\ref{ases_secrets}), and enable endpoints to validate whether this router information complies with their needs (\ref{endpoint_transparency}).

Specifically, a \textbf{router attribute} is a trait of an intra-AS or border router, such as its manufacturer or hard- and software.
A \textbf{router policy} is a predicate on a router defined in terms of its attributes.
For example ``a router is manufactured by \textit{A} or \textit{B}.''
A router policy may be lifted to a \textbf{path policy} by requiring that it holds for every router on a path.
An example of a path policy, with which the previous router policy is compliant, is ``all on-path routers are manufactured by \textit{A}.''
In \cref{sec:policy_specification}, we design a language for defining these policies.

An endpoint describes required router attributes using a router policy, which we will refer to as the endpoint's \textbf{preference policy}.
A preference policy can be used in different ways.
For example, it could be defined per-device,  per-user, per-application, or even on a per-packet basis (\ref{endpoint_fine_grained}).
Applications might even be distributed with a predefined policy.

\subsection{Trust Model}\label{sec:model:trust_model}
Which entities endpoints trust is subjective, and we therefore define our trust model from the point of view of individual endpoints.
In our model, the granularity of trust is at the level of ASes, i.e., an endpoint may trust some ASes more than others.

We assume that trusted ASes are not malicious: they follow all protocol steps correctly and they implement measures to protect their infrastructure from attacks.
However, even highly trusted ASes can make mistakes, and parts of their infrastructure may be misconfigured.
We refer to such ASes as \emph{honest-but-clumsy}.
In practice, endpoints often have existing trust relationships with some of the ASes, based on business relationships or common legislation.
Concrete examples include government offices communicating via multiple ISPs located in their country, companies affected by regulations mandating traffic not to leave specified countries~\cite{EuropeanParliament2016}, or ISPs offering policy-based path selection for an additional fee.
In order for a source endpoint to communicate with a destination endpoint in a remote AS using \system{}, there must exist at least one valid and usable path of trusted ASes.

We also assume that ASes are directly peering such that inter-AS communication does not traverse additional routers and that relevant certificates are secure, i.e., no intermediate entity, such as a certificate authority, in a certificate chain is compromised.
Furthermore, communicating endpoints are assumed to trust each other in order to agree on the same on-path router attributes.
Our model restricts the capabilities of adversaries on an endpoint-selected path to reading, modifying, injecting, and dropping packets. 

Note that we do not make any assumptions about untrusted ASes, which can exhibit arbitrary and even  actively malicious behaviour.
If \system{} relied on BGP, where traffic can be redirected by malicious off-path adversaries by  attacks such as hijacking~\cite{bgp_hijacking} or blackholing~\cite{bgp_blackholing}, endpoints would additionally need to trust a potentially large set of off-path ASes to not perform such attacks.
This observation motivates building \system{} upon SCION, which mitigates re-routing attacks by authenticating routing information using its \mbox{CP-PKI} and by enforcing correct forwarding by means of PCFS.

%% file: sections/system.tex
\section{\system{} System Description} \label{sec_system}
\System{}, based on SCION, enables ASes to announce paths along with intra-AS path policies and allows endpoints to select routes based on their own policy preferences.
\System{} satisfies the requirements of endpoints and ASes specified in \cref{sec:model:system_objectives}.
In particular, an attacker can neither change the forwarding paths and policies chosen by the endpoints nor spoof the path policies announced by the ASes.
\System{} introduces minimal processing and state overhead on routers and endpoints and is incrementally deployable.

\subsection{Overview}
\begin{figure*}[t]
    \centering
    \includegraphics[width=\textwidth]{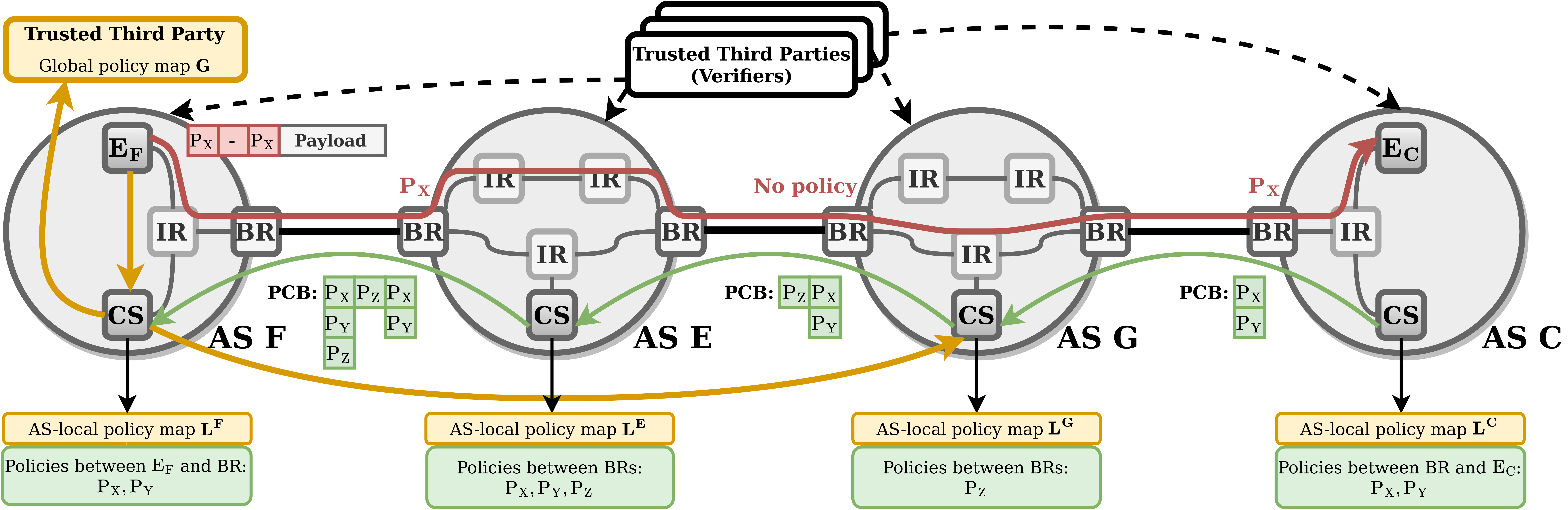}
    \caption{
    System overview.
    During beaconing, every AS adds its supported policies to the PCB (green).
    Endpoint $\text{E}_\text{F}$ fetches the corresponding path including the added policies from its control service, which resolves and caches unknown policies (orange).
    For packets destined to $\text{E}_\text{C}$, $\text{E}_\text{F}$ encodes its policy choice $\text{P}_\text{X}$ in the respective headers, except for AS G, which does not support this policy, and AS~F, which supports the policy by default (red).
    Although not all ASes support $\text{P}_\text{X}$, $\text{E}_\text{F}$ still decides to send its traffic over the path.
    Optionally, dedicated third parties verify the AS-attested policies through remote attestation.
    }\label{fig:system-overview-extended}
\end{figure*}

\System{} consists of two main processes.
In the control plane, path policies are distributed to endpoints, which compare them for compliance with their local preference policies.
In the data plane, endpoints encode the selected policies into their data packets, such that on-path ASes know how to forward them through their internal network.
A system overview is provided in \cref{fig:system-overview-extended}.

Every participating AS decides which path policies it wants to disseminate.
When originating or forwarding a SCION path construction beacon (PCB), the AS extends the PCB with path policies and AS-local identifiers, called policy indices, of internal paths that comply with these path policies.
In \cref{fig:system-overview-extended}, AS~C adds policies $\text{P}_\text{X}$ and $\text{P}_\text{Y}$, AS~G adds policy $\text{P}_\text{Z}$, and AS~E adds all three policies.
Endpoints fetch path segments (constructed from the PCBs) from the local control service and filter out paths with untrusted ASes or with path policies that do not comply with their preference policy.
If there is no path left, an endpoint can either decide not to send any traffic, or to still use one of the filtered sub-optimal paths.
The endpoint $\text{E}_\text{F}$ in \cref{fig:system-overview-extended} trusts all on-path ASes and decides to send traffic on the selected path despite one AS (AS~G) not supporting the desired policy $\text{P}_\text{X}$.
This could be a sensible decision if $\text{P}_\text{X}$ corresponds to a ``best-effort'' policy, for example, PTP support, but might not be acceptable for security-related attributes.
After selecting a set of path segments, the contained policy indices are encoded by the endpoint in a SCION packet and used by ingress border routers to forward the packet across an internal path complying with the corresponding policy.
For the first AS (AS~F in our example), the endpoint can only learn the supported policies, but it can not actively select a policy, as there is anyway only one intra-AS path to the egress border router.
Therefore, the endpoint does not need to add a policy index to the SCION header for its own AS.

A policy index does not necessarily correspond to a single intra-AS path.
The policy index encoded in the packet is evaluated when the packet arrives at the border router.
This allows the border router to choose \emph{any} intra-AS path that is compliant with the requested path policy.
Hence, intra-AS routes can be changed on-the-fly by network operators, load balancing among multiple intra-AS paths is possible, and fewer details about the intra-AS topology must be revealed.

To prevent malicious entities from negatively impacting the path selection and the endpoint privacy, our solution employs several security mechanisms of SCION.
In particular, \system{} encodes path policies in SCION PCBs and makes use of SCION’s CP-PKI for verifying signatures from the respective ASes, thereby preventing the spoofing and tampering of policies.
Moreover, using EPIC, a data plane extension for SCION, the endpoint can verify that its packet indeed traversed each on-path AS (see \cref{sec:background:epic} for  details about EPIC).
Finally, \system{} authenticates and encrypts the policy index in the packet header, providing privacy regarding an endpoint's policy preferences.

\subsection{Policy Announcement} \label{sec:system:policy_announcement}

A straightforward approach for disseminating path policies would be for ASes to encode their policy descriptions directly in the PCBs (achieving requirements \ref{endpoint_transparency} and \ref{ases_secrets}).
However, this is not space efficient since it can substantially increase the size of the PCBs for complex path policies, and therefore violate requirement \ref{ases_efficient_distribution}.
Moreover, many ASes may have similar path policies, leading to redundancy.
We therefore distinguish between globally and locally announced path policies.

\paragraph{Global Announcement}
Global policies are stored in a global append-only registry managed by a trusted entity such as ICANN~\cite{icann}.
This registry can be described as a map from a \emph{policy identifier} (PID) to the concrete policy description:
\newcommand{\globalmap}[0]{\ensuremath{\text{G}}}
\newcommand{\globaldesc}[0]{\ensuremath{\text{S}_\text{G}}}
\newcommand{\globalpol}[0]{\ensuremath{\text{P}_\text{G}}}
\begin{align*}
    \globalmap{}: \globalpol{} \rightarrow \globaldesc{} \,.
\end{align*}
Here, $\text{P}_\text{G}$ is the set of global policy identifiers and \globaldesc{} denotes the set of global path policy descriptions.
The PIDs can be short in practice---we consider a 32-bit number a reasonable representation.
The purpose of the global registry is to support globally accessible path policies used by a large fraction of ASes.
Entities can then simply refer to the PID instead of a potentially large policy description.
Entries in the registry must never be changed or removed to ensure a globally consistent view despite entities caching the registry information.

\paragraph{Local Announcement}
An AS is not restricted to only use globally defined policies, and an AS~X can define its own append-only mapping of policy identifiers:
\newcommand{\localmap}[1]{\ensuremath{\text{L}^\text{#1}}}
\newcommand{\localdesc}[1]{\ensuremath{\text{S}_\text{L}^\text{#1}}}
\newcommand{\localpol}[1]{\ensuremath{\text{P}_\text{L}^\text{#1}}}
\begin{align*}
    \localmap{X}: \localpol{X} \rightarrow \localdesc{X} \,.
\end{align*}
Here, $\text{P}_\text{L}^\text{X}$ is the set of PIDs defined by AS~X, and \localdesc{X} denotes its set of path policy descriptions.
The mapping \localmap{X} is stored at the control service of AS~X. The service is extended to also serve path policy requests based on this information.
Irrespective of whether an entity requests the policy mapping from a specific AS or from the global registry, the response must always be authenticated using a signature.
The control service caches entries from the global registry and other ASes for its local endpoints to reduce the communication overhead.

\subsection{Policy Dissemination via PCBs} \label{sec:system:policy-dissemination}

ASes disseminate their supported path policies to the endpoints by piggybacking this information on SCION PCBs.
This added information  describes which path policies are supported on specific ingress-egress interface-pairs.
Naturally, the interface pair the PCB is traversing is the most relevant; however, SCION also supports shortcut and peering links~\cite{Chuat2022a}, and therefore path policies for multiple interface-pairs can be specified.
Different path policies can be described depending on directionality, that is, an interface-pair (i,~e) need not necessarily support the same policies as the interface-pair (e,~i).
Furthermore, we allow an AS to announce path policies not only for interface-pairs, but also between an interface and an IP address range inside the AS.
This enables endpoints to also learn the policies supported by the first and last on-path AS.

\newcommand{\intf}[1]{\ensuremath{\text{IF}^\text{#1}}}
\newcommand{\ips}[1]{\ensuremath{\text{IP}^\text{#1}}}
\newcommand{\ifip}[1]{\ensuremath{\text{IFIP}^\text{#1}}}
\newcommand{\intfmap}[1]{\ensuremath{\text{I}^\text{#1}}}
\newcommand{\dpmap}[1]{\ensuremath{\text{D}^\text{#1}}}

Because endpoints have access to the global map (\globalmap{}), as well as to the local policy map of an AS~X (\localmap{X}), it suffices for the AS to only add the relevant PIDs to the PCB, instead of the full path policy descriptions. For every (global and local) PID it supports, an AS further defines a short \SI{16}{\bit}, non-zero \emph{policy index} (\ind{}{}).
The short length of the policy index is motivated by its subsequent use in the data plane, where it will be included in the data packets sent by endpoints.
The validity period of a policy index is the same as the validity period of the PCB to which it is added.
A PCB can be converted to a path segment and thus be used to send traffic usually for a duration of a few hours, after which an endpoint must get path segments from a more up-to-date PCB.
An AS can update a policy index such that it points to a different PID, but with the restriction that no two PCBs with overlapping validity periods include different policy index mappings.
Naturally, an AS must ensure that it only announces policies that it can support during the whole validity period of a PCB (\ref{endpoint_path_stability}).
However, it is free to decide how to route traffic along internal paths that satisfy the announced policies.
Some of the most prevalent routing protocols suitable for this purpose are MPLS~\cite{RFC3031} and segment routing~\cite{RFC8402}.
Based on the preceding description, we formalize the information added to a PCB by AS X in the form of two maps \intfmap{X} and \dpmap{X} as follows: 
\begin{align*}
    \intfmap{X}: \ifip{X} \rightarrow \mathcal{P} (\psi^\text{X}), \\
    \dpmap{X}: \psi^\text{X} \rightarrow \text{P}_\text{L}^\text{X} \cup \text{P}_\text{G},
\end{align*}
where
$\ifip{X} = ((\intf{X} \cup \ips{X}) \times (\intf{X} \cup \ips{X})) \setminus (\ips{X} \times \ips{X})$,
\intf{X} is the set of interfaces, \ips{X} the set of IP address ranges, and $\psi^\text{X}$ the \SI{8}{\bit} space of valid policy indices of AS X.
$\mathcal{P} (\psi^\text{X})$ denotes the power set of $\psi^\text{X}$.

\begin{figure}[t]
  \includegraphics[width=\linewidth]{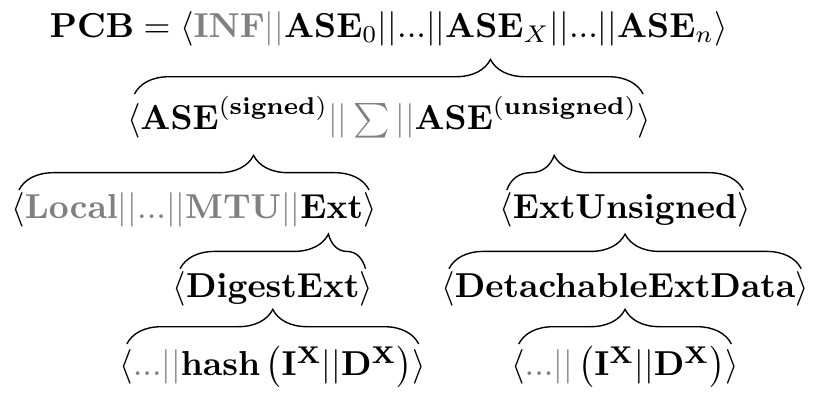}
  \caption{
  PCB format containing path policy information.
  \System{} extends the DigestExt and DetachableExtData fields of a SCION PCB by $\text{hash}(\text{I}^\text{X} || \text{D}^\text{X})$ and $\text{I}^\text{X} || \text{D}^\text{X}$, respectively.
  }\label{fig:pcb-format}
\end{figure}

To prevent manipulation by unauthorized entities, this path policy information, i.e., \intfmap{X} and \dpmap{X}, is signed by AS X (\ref{endpoint_attestation}).
\Cref{fig:pcb-format} shows the format of a PCB and how it is extended with signed policy information.
Although policy information, in the form of the two maps \intfmap{X} and \dpmap{X}, is added in the unsigned part of the PCB, this information cannot be modified, since a cryptographic hash of the map content is included in the signed part of the PCB.
Even though the two maps \intfmap{X} and \dpmap{X} can be encoded efficiently in terms of space, they can significantly increase the size of a PCB in case (i) there are many on-path ASes adding information to the PCB, (ii) many different policies are announced, or (iii) some ASes have many pairs of interfaces or IP subnets with support for policies.
Including only the hash of the maps in the signed part of the PCB allows removing the actual maps from the unsigned part of the PCB if needed (\ref{ases_efficient_distribution}).
In this case, the endpoints need to actively fetch the maps from their local AS, which fetches them from the corresponding ASes and caches them to reduce network overhead.
They can efficiently verify the authenticity of \intfmap{X} and \dpmap{X} by computing their hash and comparing it against the hash included in the PCB for AS~X.

\subsection{Endpoint Policy Selection} \label{sec_endpoint_policy_selection}

An endpoint assembles the desired end-to-end path from \mbox{up-,} core-, and down-segments constructed from the PCBs fetched from the local control service of its AS (see \cref{sec:background}).

As every PCB also contains the path policies that the corresponding ASes support, the endpoint can verify whether the corresponding segment satisfies its requirements (\ref{endpoint_transparency}).
If some of the ASes that are part of the segment are not trusted, or announce path policies that are not in accordance with the endpoints' needs, the endpoint either decides to accept such untrusted ASes or non-optimal path policies and to still use the segment, or to take advantage of SCION's path-awareness and try to find another segment that actually satisfies its preference policy.
An endpoint receives from the control service, by default, many possible segments to reach the requested destination AS, and therefore it has several possible paths and policies that it can readily choose from.

To achieve policy selection at the packet level (\ref{endpoint_fine_grained}), the endpoint communicates which policy should be satisfied to each on-path AS, except its local AS, in the form of policy indices (a policy index of zero signifies no specific desired policy).
An endpoint can only select a single policy index per on-path AS; an AS can announce local path policies that are conjunctions of other path policies.
Upon reception of a data packet, an ingress border router parses the policy index and maps it to an internal path satisfying the corresponding path policy.
This mapping, which can be implemented efficiently using a hash table (\ref{ases_efficient_forwarding}), adds overhead to the processing time of every packet, which is however negligible  even for ASes supporting hundreds of policies (\cref{sec:implementation}).
In particular, its scalability does not depend on the number of endpoints (hosts, applications, or flows) actively sending traffic.

The codomain of the mapping and any further actions executed on a packet depend on the intra-AS routing protocol in use.
In MPLS, for example, a policy index is mapped to a set of labels, each corresponding to a single label switched path (LSP). The border router places the selected label in the packet header, which is later again removed at the egress border router.
To avoid packet reordering when multiple intra-AS routes are available, the border router can use hashing to always map packets of the same flow to the same intra-AS route.
The intra-domain routing protocol and the soft- and hardware of internal routers need not be modified.

If the packet contains a policy index that is not supported, the border router sends back an authenticated control message to the endpoint.
Because a border router only receives policy-enabled packets that are source-authenticated (\cref{sec_securing_policy_indices}), control messages can not be abused by an attacker to launch reflection DoS attacks against other endpoints.

Upon request, the control service also provides policy information for the intra-AS path between two endpoints of the same AS.
Because intra-AS communication does not use the SCION protocol, an endpoint cannot select the preferred policy; however, it learns from the control service which path policies are guaranteed to be satisfied when sending traffic within the AS to the other endpoint, including their validity period.
By using short periods, the AS can propagate routing configuration updates quickly to the endpoints.
The path policies supported on the route between the endpoint and the egress border router are part of the PCB.

An endpoint's preference policy can be set in various ways, depending on the concrete use case.
There can be a single policy for the entire device, or an application-specific policy, e.g., for an application sending sensitive corporate data.
Per-application policies are set before the connection establishment is initiated, similar to TAPS~\cite{RFC8923}, which provides fine-grained control over transport protocol parameters to applications.
Finally, different policies can be set for individual packet flows or even on a per-packet basis, but requires modifying the endpoint's network stack.

\subsection{Securing Policy Indices} \label{sec_securing_policy_indices}

The policy indices in data packets must be authenticated, otherwise malicious entities could modify them and thus violate the endpoint's preference policy.
Furthermore, indices in the packets should be hidden, such that an index is only accessible by the endpoint and the on-path AS that announced it.

To achieve authenticity and secrecy of the policy indices, we leverage the fast key derivation provided by DRKey~\cite{DRKey,PISKES} and a modified version of EPIC~\cite{legner2020epic}.
In EPIC, a source host includes short per-packet hop validation fields (HVFs) in the packet header, which are subsequently verified by the on-path border routers to assess the authenticity of the packet source.
Through its path validation feature, EPIC furthermore allows the source and destination endpoints to verify for each packet that it has indeed traversed the ASes of the correct, i.e., the previously selected, path. Therefore, EPIC augments \system{} to satisfy requirement \ref{endpoint_path_validation}.
Background on DRKey and EPIC can be found in \cref{sec:background:drkey,sec:background:epic} and our modification to EPIC that also supports authenticity and confidentiality of the policy indices is described in \cref{sec:appendix:computation}.
Through these modifications, \system{} also satisfies requirement \ref{endpoint_privacy}.

With this solution, unauthorized entities can neither infer nor modify the policies chosen by endpoints. Moreover, it is not possible to detect whether an endpoint chooses a policy at all, as an \emph{encrypted} zero and non-zero index are indistinguishable.
Irrespective of its value, a policy index is always \SI{16}{bits}---also in its encrypted form, which is achieved by one-time pad encryption.
Due to the per-packet cryptographic operations, the encrypted policy indices and hop validation fields change for each packet sent, and hence packets with the same policy index also cannot be distinguished.

\subsection{Router Attestation}\label{sec:router-attestation}
Remote attestation on routers that are equipped with a trusted platform module (TPM) can ensure that the device is running the intended software and is not compromised---and is thus considered ``trustworthy''~\cite{rats}.
If every single device on the forwarding path is verified to be trustworthy, then the complete forwarding path is considered trustworthy.

Router attestation can improve \system{} by strengthening the path policy claims of an AS (\ref{endpoint_attestation}).
This is the case when the attestation result provides enough information to conclude that the routers are compliant with an endpoint's preference policy.
This information can be used to augment the trust placed in an AS, as an orthogonal measure to the signed path policies in the PCBs.
While such router attestation within an AS cannot generate a proof that traffic is actually routed over these devices, it can be used to determine the veracity of path policies claimed in PCBs by proving that a path-policy-compliant intra-AS path \textit{exists}.
If an AS is trustworthy, it will comply and send traffic over these policy-compliant routers.

An AS can release the attestation results to a (trusted) third party that then verifies the existence of a path-policy-compliant intra-AS path on behalf of the users.
Thereby, the endpoint does not receive the actual attestation result. The third party instead returns a signed statement stating that all path policies were validated based on the AS' attestation results.
Alternatively, attestation results could be published on a public append-only log.
However, publicly sharing attestation results can be problematic as this might reveal sensitive information about the AS-internal network and is thus only done if approved by the respective AS (\ref{ases_secrets}).

%% file: sections/specification.tex
\section{Policy Specification} \label{sec:policy_specification}

In \system{}, desirable router attributes are specified by router policies, which are predicates on a router's configuration.
When this predicate is satisfied for a router's configuration, the policy is said to comply with the configuration.
By specifying policies as formulas in a sorted first-order-logic (FOL) referencing relevant router attributes, users can define arbitrarily complex policies.
The system remains extensible, since this language can be enriched with new router attributes, should they become relevant in the future.

\begin{table}[t]
  \centering
  \caption{The different sorts and their respective carrier sets.}
  \begin{tabular}{ccc}
    \toprule
    Sort & Carrier Set & Type \\
    \midrule
    Manufacturer ($M$) & $\mathcal{M}$ & \makecell{private enterprise\\number (int)~\cite{iana-enterprise-numbers}} \\\midrule
    \makecell{Software\\Component ($C$)} & $\mathcal{C}$ & unique ID \\\midrule
    Tag ($T$) & $\mathcal{T}$ & string \\\midrule
    Tag Issuer ($I$) & $\mathfrak{I}$ & URI (string) \\\midrule
    Name ($N$) & $\mathcal{N}$ & string \\\midrule
    Version ($V$) & $\mathcal{V}$ & \makecell{version\\scheme (string)} \\\midrule
    Router ($R$) & $\mathcal{R}$ & unique ID \\\midrule
    Path ($P$) & $\mathcal{P}$ & unique ID \\
    \bottomrule
  \end{tabular}
  \label{tab:sorts}
\end{table}

\subsection{Syntax and Semantics}\label{sec:policy-syntax}\label{sec:system:policy:semantics}
Policies are encoded in a sorted FOL with equality as formulas with free variables.
In addition to the general FOL syntax, given in \cref{sec:appendix:policy-syntax}, we define the sorts $M, C, T, I, N, V, R$, and $P$ (see \cref{tab:sorts}), the function symbols \texttt{tag} (of type $C \rightarrow T$), \texttt{issuer} (of type $T \rightarrow I$), \texttt{name} (of type $C \rightarrow N$), \texttt{version} (of type $C \rightarrow V$), \texttt{manufacturer} (of type $R \rightarrow M$), the predicates \texttt{onPath} (of type $P \times R$) and \texttt{software} (of type $R \times C$), and the comparison predicates $<$, $\leq$, $\geq$, and $>$ (of type $V \times V$).

A path consists of a set of routers (ignoring their order) and each router is described by its router setup.
A router setup specifies the router's manufacturer and the router's software stack, which consists of software components.
A software component is specified by a name, a version, and a globally unique tag, issued by a tag issuer.
\Cref{sec:appendix:semantics} contains a more detailed description of the semantics and interpretation.

\subsection{Policy Definition}\label{sec:system:policy:definition}
Both user preference policies (\ref{endpoint_transparency}) and path policies (\ref{ases_secrets}) are defined as router policies.
However, they are used for different purposes, since a user preference policy defines all acceptable router setup(s) of a user, whereas a path policy defines all possible router setups on a path.

We define a router policy as an open formula of the form $Pol(r)$, with one free variable $r$ of sort $R$.
A router policy accepts a router setup iff the formula $Pol(r)$ is satisfied ($\models$) in an interpretation $\mathcal{I}$ with respect to an assignment $\alpha$, which assigns a router to the free variable $r$, where $\alpha: \{r\} \rightarrow \mathcal{R}$.
\[ \mathcal{I}, \alpha \models Pol(r) \]
A \textbf{path policy} ($PathPol_{p}$) advertised for a path $p$ must accept the setups of all on-path routers ($R$):
\[ \forall r' \in \mathcal{R}: \texttt{onPath}(p, r') \rightarrow PathPol_p(r') \]
An example of a path policy covering routers with a specific version $v$ of a software $s$ issued by issuer $i$ and a manufacturer $m$, using $\texttt{manu}()$ as a shorthand for $\texttt{manufacturer}()$, is
\begin{equation*}
  \begin{aligned}
    PathPol_p(r)~:= ~\texttt{manu}(r)=m \wedge \exists c \in \mathcal{C}: \texttt{software}(r,c)& \\
    \wedge ~\texttt{name}(c) = s \wedge \texttt{issuer}(\texttt{tag}(c)) = i \wedge \texttt{version}(c) = v&
  \end{aligned}
\end{equation*}
A \textbf{user preference policy} $PrefPol_{u}(r)$ describes acceptable on-path router setups of user $u$.
The user automatically decides to accept or reject a path $p$ based on its preference policy and the path policy $PathPol_{p}(r)$.
If $PathPol_{p}(r)$ is contained in $PrefPol_{u}(r)$, then $u$ accepts $p$, otherwise $u$ rejects $p$.

An example of a user preference policy that requires routers from either manufacturer $m_1$ or $m_2$, where a critical software ($s_\text{crit}$) was recently updated (at least version $v_\text{min}$), is
\begin{equation*}
  \begin{aligned}
    PrefPol_u(r)~&:= ~(\texttt{manu}(r)=m_1 \vee \texttt{manu}(r)=m_2)\\
    &\wedge \forall c \in \mathcal{C}: (\texttt{software}(r,c) \wedge \texttt{name}(c) = s_{\text{crit}}\\
    &\wedge \texttt{issuer}(\texttt{tag}(c)) = i) \rightarrow \texttt{version}(c) \geq v_{\text{min}}
  \end{aligned}
\end{equation*} 

Determining whether a path policy is contained in a user’s preference policy is equivalent to determining query containment.
Note, however, that query containment can not always be evaluated efficiently.
In general, query containment in FOL is known to be undecidable~\cite{Mostowski1950-IOA-10}.
Even in a restricted subset of FOL, namely FOL formulas consisting of a disjunction of conjunctive queries, query containment is NP hard~\cite{Chandra1977}.
We expect that most path policies are relatively simple,  and that user preference policies are not overly complex either.
Additionally, the outcome of complex query containment evaluations can be cached locally with minimal overhead as long as the policies are in use.
As discussed in \cref{sec:system:policy-dissemination}, path policies are not shared directly in PCBs, but instead fetched on demand by the endpoint from the global policy repository or the local repository of the respective AS, while preference policies are defined locally and never shared with other entities.

%% file: sections/implementation.tex
\section{Implementation and Evaluation}\label{sec:implementation}
In this section, we demonstrate that \system{} works efficiently in practice. Parties interested in this research can review our implementation and modify it according to their needs with minimal effort. Because our per-packet authentication and encryption operations pose an additional overhead to the sender and the on-path border routers, we also implement and evaluate a high-speed version of those components.
For a path consisting of four ASes and packets with payloads of \SI{1000}{\byte}, the sender and the router achieve \SI{3.29}{Gbps} and \SI{14.62}{Gbps} per processing core, respectively.
The performance increases linearly with the number of cores.
We also experimentally demonstrate the feasibility of TPM-based remote router attestation for use in \system{}.

\subsection{Deployment in SCIONLab}\label{sec:evaluation:scionlab}
We implement \system{} in SCIONLab~\cite{scionlab,scionlabpaper}, a global network testbed that enables research and experimentation with SCION. Users can join SCIONLab through one or more SCION ASes.
While some ASes are connected through dedicated links, the majority of inter-AS links are IP overlays.
For our purposes, we set up two ASes and connect them to SCIONLab at two different access points, each in a different country.
The first AS~P is modified such that the control service extends beacons with its set of policy indices. We also adapt its border routers, such that they simulate intra-AS routes with different latencies.
Our second AS~S contains a SCION-enabled endpoint, which we modify such that it prompts the user to not only select a path, but also one policy per policy-aware on-path AS. The endpoint then creates data packets carrying the chosen policies.
We implement this by means of a hop-by-hop extension\cite{hopbyhopheader}, which is ignored by all on-path ASes that do not support \system{}, but parsed and handled by AS~S. Thus no changes are necessary in SCION ASes that are outside of our control.
For the evaluation, we measure the round-trip time of control message packets for different policies selected by the user. 
We took precautions to not harm any real network infrastructure: Our experiments were directed at specific benchmarks and small in scale.
With one packet per second, our evaluation adds negligible overhead, and thus does not negatively impact other parties.
We obtained permission from the SCIONLab operators to conduct our measurements.
As an alternative to this setup, we also provide a system for simulation-based evaluation allowing the execution of arbitrary Internet topologies on a single machine.\footnote{\url{https://github.com/mawyss/scion/tree/policy_routing}}
This supports the fast verification of our results, and low-effort modifications to test new ideas.

\subsection{High-Speed Implementation}
To secure the policy indices in the packet header, we leverage DRKey~\cite{DRKey,PISKES}, a protocol to efficiently derive symmetric keys, and EPIC~\cite{legner2020epic}, a system for fast per-packet source authentication and path validation in SCION, modified to ensure the authenticity and confidentiality of the policy indices.
To demonstrate that our modifications to EPIC still allow for high-speed packet processing, we implement and evaluate the components affected by our changes, namely the source endpoint and the border routers (the EPIC protocol running on the destination endpoint is not modified).
Such an evaluation is important because we encrypt, authenticate, and verify the policy-indices on a per-packet basis, which must be fast enough for practical deployments.
Our modifications are described in \cref{sec:appendix:computation}.
We implement the source endpoint mechanism to create per-packet validation fields and the corresponding verification procedure at the border router using Intel DPDK~\cite{dpdk}, instantiate MAC computations with AES-CBC-MAC, and rely on Intel AES-NI~\cite{AESni} to speed up AES block cipher computations.
Our testbed consists of a commodity machine (Intel Xeon, 2.1 GHz), which executes our implementation that is to be evaluated, and a Spirent SPT-N4U, which serves both as bandwidth generator (when evaluating the border router) and bandwidth monitor (when evaluating the source endpoint). Both machines are connected by four \SI{40}{Gbps} bidirectional Ethernet links. We evaluate both implementations (endpoint, border router) individually as they never run at the same time on the commodity machine.

%% file: sections/evaluation.tex
\subsection{Evaluation}\label{sec:evaluation}
We now present our results of the RTT measurements in SCIONLab and the high-speed evaluation of the policy-enabled sender and router.
Additionally, we evaluate the control-plane overhead and TPM-based remote attestation.

\begin{figure}[t]
    \input{figures/scionlab_latency}
    \caption{Round-trip time (RTT) measurements for different selected policies over a period of \SI{60}{\s} conducted at an endpoint in source AS~S for a 5-hop path going through on-path AS~P.
      A policy index ($\ind{}{}$) of zero and one corresponds to the standard and low-latency intra-AS route, respectively.}
    \label{fig:scionlab_latency}
\end{figure}
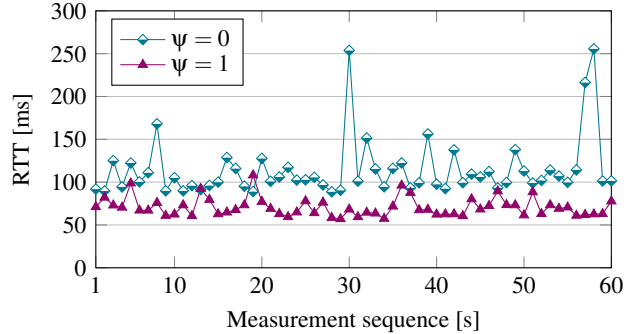

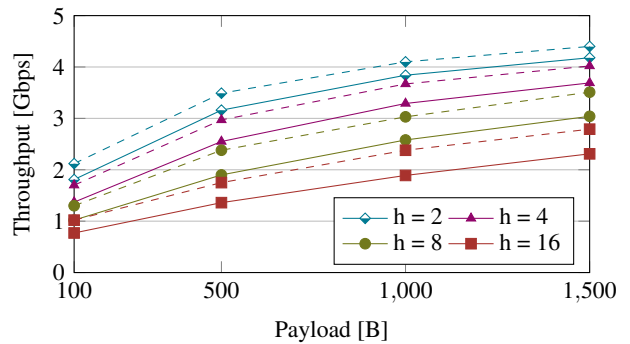
\begin{figure}[t]
    \input{figures/policy_sender}
    \caption{Source endpoint packet generation performance for different number of AS-level hops (h) and payloads, using one CPU core.
    Dashed lines correspond to the original EPIC without support for policies.}
    \label{fig:policy_sender}
\end{figure}

\begin{figure}[t]
    \input{figures/policy_router}
    \caption{Border router packet forwarding performance for various payload sizes and different number of CPU cores.
    Dashed lines have the same meaning as in \cref{fig:policy_sender}.
    }
    \label{fig:policy_router}
\end{figure}
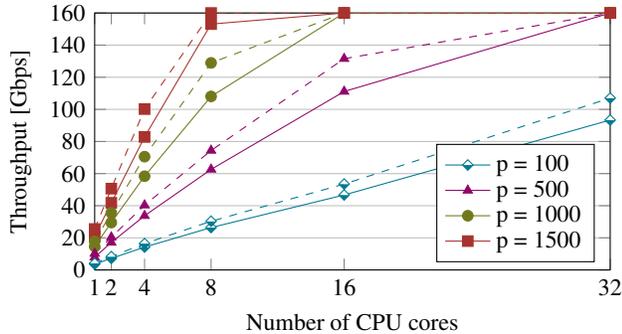

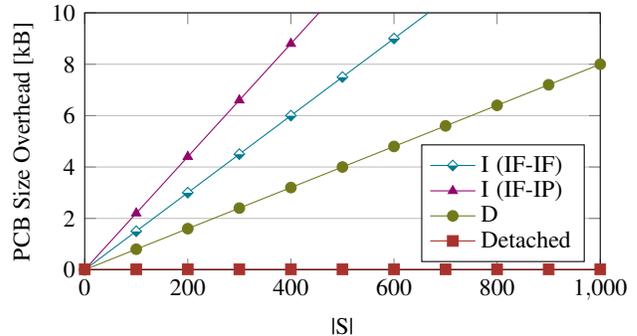
\begin{figure}[t]
    \input{figures/pcb_size}
    \caption{
    Per-AS PCB size overhead in kilobytes introduced for the maps $\intfmap{}$, $\dpmap{}$ (see \cref{sec:system:policy-dissemination}), and the detached extension. The x-axis is the number of elements in set S, which is a subset of either $\intf{}\times\intf{}$,  $(\intf{}\times\ips{})\cup(\ips{}\times\intf{})$, or $\psi^\text{X}$.
    The number of announced policy indices per IF-IF or IF-IP pair is five.
    }
    \label{fig:pcb_size}
\end{figure}

\begin{figure}[t]
    \centering
    \includegraphics{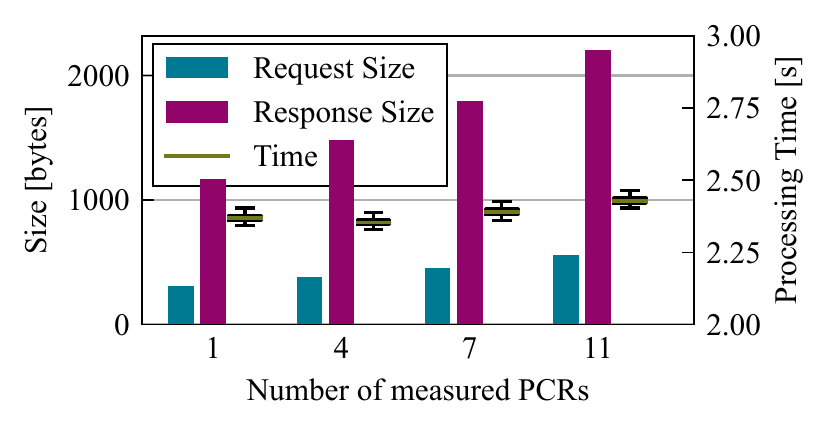} 
    \caption{Request and response sizes and processing times for fetching various numbers of PCR measurements.
    }
    \label{fig:fetch_fetch_tpm_quotes}
\end{figure}

\begin{itemize}[leftmargin=0pt,label=]
    \setlength\itemsep{0pt}
    \item \textbf{RTT Measurements}
    \hspace{3pt}
    In our SCIONLab setup, AS~$P$ announces two policy indices, where index zero denotes a standard intra-AS path, and index one a low-latency path.
    We choose this type of policy because, in contrast to router attributes such as manufacturer or software version, latency is a characteristic directly measurable by the communicating endpoints.
    The results of our experiments performed by an endpoint in AS~$S$ are shown in \cref{fig:scionlab_latency}.
    While packets traversing the standard intra-AS route of on-path AS~$P$ travel for \SI{114}{\ms} on average and suffer from high jitter, packets carrying a policy index equal to one take on average \SI{70}{\ms} per round-trip and never exceed \SI{110}{\ms}.
    \item \textbf{Endpoint Evaluation}
    \hspace{3pt}
    \cref{fig:policy_sender} shows the evaluation results for a single CPU core of the source endpoint.
    When sending packets with a payload of \SI{1000}{\byte} on a path consisting of four AS-level hops (the average path length in today's Internet\cite{distance_metrics,internet_modeler,average_path}), the endpoint can generate traffic at a rate of \SI{3.29}{Gbps}.
    As expected, the throughput increases with the size of the packet payload and decreases with the number of on-path ASes, where the latter is caused by the per-AS encryption and authentication operations.
    Although one core of the source endpoint already provides particularly high performance, the results can be scaled linearly with respect to the number of CPU cores dedicated to the generation of policy-enabled traffic.
    For payloads of \SI{1000}{\byte}, \system{} is \num{7}-\SI{20}{\%} slower compared to the original version of EPIC, depending on the number of ASes on the path.
    Because the endpoint can still generate traffic at gigabit rates even on a single core, this overhead does not impact its practicality.
    \item \textbf{Router Evaluation}
    \hspace{3pt}
    The results of the border router evaluation are given in \cref{fig:policy_router}.
    The router's forwarding performance depends on the size of the packet payload and the number of CPU cores used, but is independent of the number of on-path ASes, because the router only validates the packet header fields dedicated to its own AS.
    For instance, using \num{16} cores and processing packets with a payload of \SI{1000}{\byte}, the router can handle the full capacities of all its links, achieving efficient forwarding (\ref{ases_efficient_forwarding}) with a throughput of \SI{160}{Gbps}.
    
    From fine-grained timing measurements we discovered that the router's per-packet processing overhead is only slightly higher than the standard EPIC router operations.
    In particular, checking the authenticity of the encrypted policy index takes additional \SI{15}{\ns}, decrypting the encrypted policy index amounts to \SI{52}{\ns}, and looking up an intra-AS path corresponding to the policy index increases the processing overhead by \SI{35}{\ns} when \num{1000} indices are supported in total.
    This corresponds to a decrease in throughput of \textasciitilde\SI{17}{\%} compared to the original version of EPIC.
    Due to EPIC's efficient design, this relative overhead is noticeable. However, in real-world deployments, this is unproblematic because (i) the absolute number of cores needed to compensate for this overhead is still low and (ii) the linear scalability of the border router still allows ASes to deploy \system{} in various environments with different network capacities.
    The router throughput and latency are independent of the number of communicating endpoints or their selected policies, since for every packet the same operations are executed. The number of policy indices stored at a router only affects intra-AS path lookup, which is efficient due to its implementation based on a hash table.
    \item \textbf{PCB Size Overhead}
    \hspace{3pt}
    The communication overhead imposed on the PCB by a single AS is shown in \cref{fig:pcb_size}, where policy information is encoded using protocol buffers~\cite{protobuf}.
    For a core AS announcing \num{500} interface-pairs with five policies each, with a total of \num{100} supported policies, the overhead is \SI{7.5}{\kilo\byte} for the map \intfmap{} plus \SI{0.8}{\kilo\byte} for the map \dpmap{}.
    If the policy information is detached, there is a overhead of \SI{18}{\byte}.
    \item \textbf{Remote Attestation}
    \hspace{3pt}
    Through measurements on a CISCO NCS 540 device~\cite{ncs540}, i.e., a router with TPM support, we show the feasibility of the remote attestation proposal from \cref{sec:router-attestation}.
    \Cref{fig:fetch_fetch_tpm_quotes} shows the request/response size and processing time overhead to fetch Platform Configuration Register (PCR) measurements from a single router.
    These measurements are, among others, used to infer which software is running on this router.
    The request and response sizes scale linearly with the number of measured PCRs,
    but never exceed \SI{2.2}{\kilo\byte}.
    Retrieving PCRs takes around \SI{2.4}{\second} and thus policy-violating routers can quickly be detected.
    A verifier can validate the trustworthiness of a network in a few seconds by attesting multiple routers in parallel.
    To prevent abuse, an AS enabling remote attestation must explicitly authorize verifiers to use that feature.
    The AS might implement additional security measures such as rate-limiting of requests.
\end{itemize}

%% file: figures/scionlab_latency.tex
\begin{tikzpicture}
	
	\begin{axis}
	[
	font=\small,
	xlabel={Measurement sequence [\si{\s}]},
	ylabel={RTT [\si{\ms}]},
	legend entries={$\ind{}{} = 0$, $\ind{}{} = 1$},
	xmin=1, xmax=60,
	ymin=0, ymax=300,
	xtick={1, 10, 20, 30, 40, 50, 60},
	ytick={0, 50, 100, 150, 200, 250, 300},
	ymajorgrids,
	height=5cm, width=\axisdefaultwidth,
	legend style={cells={anchor=west}, row sep=-2pt},
	legend pos=north west,
	cycle list name=eval list fabridcyclelist,
	]
	
	\addplot
	table {%
    	1	 91.871
        2    89.636
        3    125.161
        4    94.377
        5    122.316
        6    100.265
        7    110.989
        8    168.084
        9    89.844
        10    105.147
        11    89.852
        12    95.898
        13    91.255
        14    96.066
        15    100.015
        16    128.736
        17    115.94
        18    95.202
        19    89.063
        20    127.687
        21    100.935
        22    105.972
        23    117.305
        24    102.395
        25    102.697
        26    105.62
        27    96.678
        28    88.726
        29    90.495
        30    253.848
        31    101.109
        32    151.543
        33    115.411
        34    94.55
        35    116.06
        36    122.357
        37    93.326
        38    99.238
        39    156.238
        40    97.51
        41    92.447
        42    137.659
        43    99.365
        44    109.396
        45    106.063
        46    112.413
        47    93.182
        48    99.468
        49    138.019
        50    112.962
        51    98.982
        52    102.012
        53    114.079
        54    107.093
        55    99.805
        56    114.28
        57    216.391
        58    255.607
        59    101.336
        60    101.495
	};

	\addplot
	table {%
		1    71.032
        2    82.084
        3    72.961
        4    70.396
        5    98.685
        6    67.04
        7    66.992
        8    75.878
        9    60.831
        10    62.381
        11    73.193
        12    60.607
        13    92.227
        14    79.515
        15    62.769
        16    64.708
        17    67.644
        18    73.213
        19    108.139
        20    77.166
        21    69.014
        22    62.96
        23    59.378
        24    64.824
        25    77.92
        26    64.009
        27    76.273
        28    58.413
        29    57.001
        30    68.193
        31    59.356
        32    64.56
        33    63.554
        34    57.377
        35    71.703
        36    96.162
        37    87.632
        38    67.442
        39    67.936
        40    62.316
        41    62.568
        42    62.703
        43    60.539
        44    80.23
        45    68.279
        46    72.332
        47    89.506
        48    73.395
        49    73.184
        50    61.386
        51    88.4
        52    63.004
        53    73.241
        54    69.107
        55    70.618
        56    61.006
        57    61.888
        58    62.56
        59    62.999
        60    77.703
	};

	\end{axis}
\end{tikzpicture}

%% file: figures/policy_sender.tex
\begin{tikzpicture}
	
	\begin{axis}
	[
	font=\small,
	xlabel={Payload [\si{\byte}]},
	ylabel={Throughput [\si{Gbps}]},
	legend entries={h = 2, h = 4, h = 8, h = 16},
	xmin=100, xmax=1500,
	ymin=0, ymax=5,
	xtick={100, 500, 1000, 1500},
	ytick={0, 1, 2, 3, 4, 5},
	ymajorgrids,
	height=5cm, width=\axisdefaultwidth,
	legend style={cells={anchor=west}, row sep=-2pt},
	legend pos=south east,
	legend columns=2,
	cycle list name=eval list fabridcyclelist,
	]
	
	\addplot
	table {%
    	100	    1.81
        500     3.16
        1000    3.84
        1500    4.18
	};
	
	\addplot
	table {%
    	100	    1.37
        500     2.55
        1000    3.29
        1500    3.69
	};
	
	\addplot
	table {%
    	100	    1.02
        500     1.90
        1000    2.58
        1500    3.04
	};
	
	\addplot
	table {%
    	100	    0.77
        500     1.36
        1000    1.89
        1500    2.31
	};
	
	\addplot
	table {%
    	100	    2.12
        500     3.49
        1000    4.10
        1500    4.40
	};
	
	\addplot
	table {%
    	100	    1.70
        500     2.97
        1000    3.67
        1500    4.02
	};
	
	\addplot
	table {%
    	100	    1.30
        500     2.38
        1000    3.03
        1500    3.51
	};
	
	\addplot
	table {%
    	100	    1.02
        500     1.75
        1000    2.38
        1500    2.79
	};

	\end{axis}
\end{tikzpicture}

%% file: figures/policy_router.tex
\begin{tikzpicture}
	
	\begin{axis}
	[
	font=\small,
	xlabel={Number of CPU cores},
	ylabel={Throughput [\si{Gbps}]},
	legend entries={p = 100, p = 500, p = 1000, p = 1500},
	xmin=1, xmax=32,
	ymin=0, ymax=160,
	xtick={1, 2, 4, 8, 16, 32},
	ytick={0, 20, 40, 60, 80, 100, 120, 140, 160},
	ymajorgrids,
	height=5cm, width=\axisdefaultwidth,
	legend style={cells={anchor=west}, row sep=-2pt},
	legend pos=south east,
	cycle list name=eval list fabridcyclelist,
	]
	
	\addplot
	table {%
    	1   3.61
        2   7.19
        4   14.22
        8   26.39
        16  46.69
        32  93.34
	};
	
	\addplot
	table {%
    	1   7.91
        2   17.12
        4   33.75
        8   62.60
        16  111.12
        32  160
	};
	
	\addplot
	table {%
    	1   14.62
        2   29.52
        4   58.37
        8   108.05
        16  160
        32  160
	};
	
	\addplot
	table {%
    	1   20.90
        2   41.80
        4   82.84
        8   153.11
        16  160
        32  160
	};
	
	\addplot
	table {%
    	1   4.16
        2   8.31
        4   16.42
        8   30.33
        16  53.48
        32  107.02
	};
	
	\addplot
	table {%
    	1   10.23
        2   20.39
        4   40.31
        8   74.38
        16  131.55
        32  160
	};
	
	\addplot
	table {%
    	1   17.86
        2   35.55
        4   70.54
        8   128.95
        16  160
        32  160
	};
	
	\addplot
	table {%
    	1   25.51
        2   50.65
        4   100.20
        8   160
        16  160
        32  160
	};
	
	\end{axis}
\end{tikzpicture}

%% file: figures/pcb_size.tex
\begin{tikzpicture}
	
	\begin{axis}
	[
	font=\small,
	xlabel={|S|},
	ylabel={PCB Size Overhead [\si{\kilo\byte}]},
	legend entries={\intfmap{} (IF-IF), \intfmap{} (IF-IP), \dpmap{}, Detached},
	xmin=0, xmax=1000,
	ymin=0, ymax=10,
	ymajorgrids,
	height=5cm, width=\axisdefaultwidth,
	legend style={cells={anchor=west}, row sep=-2pt},
	legend pos=south east,
	cycle list name=eval list fabridcyclelist,
	]
	
\addplot
table {%
0	0.000
100	1.500
200	3.000
300	4.500
400	6.000
500	7.500
600	9.000
700	10.500
800	12.000
900	13.500
1000	15.000
};

\addplot
table {%
0	0.000
100	2.200
200	4.400
300	6.600
400	8.800
500	11.000
600	13.200
700	15.400
800	17.600
900	19.800
1000	22.000
};

\addplot
table {%
0	0.000
100	0.800
200	1.600
300	2.400
400	3.200
500	4.000
600	4.800
700	5.600
800	6.400
900	7.200
1000	8.000
};

\addplot
table {%
0	0.018
100	0.018
200	0.018
300	0.018
400	0.018
500	0.018
600	0.018
700	0.018
800	0.018
900	0.018
1000	0.018
};

\end{axis}
\end{tikzpicture}

%% file: sections/analysis.tex
\section{Security and Scalability Analysis} \label{sec:security}
We now analyze \system{}'s security and scalability.

\subsection{Security}\label{sec:analysis:security}
We first discuss how \system{} satisfies the security requirements of endpoints and ASes (\ref{endpoint_attestation}, \ref{endpoint_privacy}, \ref{endpoint_path_validation}, \ref{ases_secrets}, \ref{ases_ddos_protection}), and then show that the Realistic Objective from \cref{sec:model:system_objectives} is achieved with respect to the trust model described in \cref{sec:model:trust_model}.

\begin{itemize}[leftmargin=0pt,label=]
    \setlength\itemsep{0pt}
    \item \textbf{Attested Claims (\ref{endpoint_attestation})}
    \hspace{3pt}
    Claims about the mapping of interface or IP pairs to policy indices (\intfmap{X}) and the mapping of policy indices to policy identifiers (\dpmap{X}) are attested by the respective AS through a signature contained in the PCB.
    When requesting the mapping from local policy identifiers to FOL policy descriptions of some AS~X (\localmap{X}), a signature over this mapping is included in the response.
    Similarly, the mapping from global policy identifiers to FOL policy descriptions (\globalmap{}) is signed by the trusted third party.
    The use of signatures to authenticate the policy information ensures the authenticity and non-repudiation of policy information, which can be confirmed by every entity possessing the corresponding public keys.
    Furthermore, \localmap{X} and \globalmap{} are append-only, so any modifications to existing entries can trivially be detected.
    Finally, claims about device properties are verified by trusted third parties through remote attestation (see \cref{sec:router-attestation}).
    Those parties vouch for the correctness of the information extracted from the attestation results and published to ASes and endpoints.
    \item \textbf{Secrecy and Authenticity (\ref{endpoint_privacy})}
    \hspace{3pt}
    The source of the packet, i.e., the three-tuple consisting of ISD, AS, and IP address, is authenticated by EPIC based on symmetric keys derived by DRKey.
    Our customized adaption of EPIC further ensures the authenticity and confidentiality of the policy indices stored in the header of data packets.
    This prevents on-path adversaries from tampering with the selected policies.
    The default policy index of zero, which indicates no preference regarding any path policies, is indistinguishable from any other policy index in their encrypted form.
    Moreover, a policy index is encrypted on a per-packet basis, and therefore an adversary can not infer whether any two packets carry the same policy indices.
    \item \textbf{Trade Secrets (\ref{ases_secrets})}
    \System{} also preserves an AS provider's trade secrets, as the kind of network information  published can be chosen by the ASes individually and at fine granularity. Thus, any undesired leakage of sensitive topology information can always be prevented.
    Even if, for example, an AS announces a policy corresponding to recently patched devices, an attacker cannot infer that traffic sent over paths with different policies will be sent over devices with known vulnerabilities. 
    However, it is the provider's responsibility to prevent unintended leakage.
    \item \textbf{Path Validation (\ref{endpoint_path_validation})}
    \hspace{3pt}
    Path validation provided by EPIC allows an endpoint to verify for every packet sent, whether it has actually traversed all desired on-path ASes.
    This prevents the violation of the endpoint's forwarding directives by misconfigured entities that are part of our trust model, i.e., honest-but-clumsy ASes.
    EPIC achieves non-probabilistic guarantees, while significantly outperforming other path validation protocols in terms of communication overhead~\cite{legner2020epic}.
    \item \textbf{DDoS Protection (\ref{ases_ddos_protection})}
    \hspace{3pt}
    \System{} does not force ASes to announce intra-AS paths directly, as this could be abused for targeted DDoS attacks.
    Instead, ASes only publish path policies, which can be mapped to multiple intra-AS paths, allowing for better distribution of ingress traffic.
    Because the source and policy index of every packet are authenticated, an on-path AS can implement further measures to protect against DDoS attacks such as rate-limiting based on the packet's source and even the selected policy index.
    \item \textbf{Summary (Realistic Objective)}
    \hspace{3pt}
    To show that \system{} achieves the Realistic Objective from \cref{sec:model:system_objectives} given our trust model, we assume that a path consisting of trusted ASes exists between an endpoint and its communication peer.
    If this is not the case, the endpoint does not send any traffic, and the objective is trivially achieved.
    \\
    Because the policy claims are all attested by the respective ASes, the endpoint can be sure that those claims have not been modified by unauthorized parties.
    However, according to our trust model, trusted ASes are considered honest-but-clumsy.
    An AS can be potentially misconfigured to (i) announce wrong policies in the PCBs, to (ii) forward packets arriving on the ingress interface along an internal path violating the specified policies, or to (iii) forward packets on a different inter-domain path than defined in the PCFS.
    \\
    The first misconfiguration alone is unproblematic, since an ingress border router will not forward traffic carrying unsupported policies.
    To mitigate the combination of the first two misconfigurations, the trusted third party verifiers detect through remote attestation that the announced policies are indeed implemented.
    Lastly, the endpoint can verify through path validation that the selected forwarding path is indeed followed by all on-path ASes.
    Therefore, we conclude that traffic is only sent along routers satisfying the endpoint-selected policies, i.e., routers with acceptable attributes, hence \system{} indeed achieves the Realistic Objective from \ref{sec:model:system_objectives}.
\end{itemize}

\subsection{Scalability}\label{sec:analysis:scalability}
\System{} proposes modifications to a global, inter-domain routing infrastructure and must thus work efficiently at scale (\ref{ases_efficient_distribution} and \ref{ases_efficient_forwarding}).
Internet-wide scalability of the underlying SCION architecture and EPIC has been analyzed in previous work~\cite{Kraehenbuehl2021,legner2020epic}.
We therefore discuss only the \emph{additional} processing, network, and state overhead of \system{}.

\system{}'s most critical component for scalability is the data plane of endpoints and border routers.
\Cref{sec:evaluation} shows the low processing overhead compared to EPIC despite using per-hop encrypted and authenticated policy indices in data-plane packets.
The packet header overhead for data-plane packets is linear with the AS--path-length (\SI{16}{\bit} per AS), which has little effect on the goodput.
Finally, border routers need to store the policy index to intra-AS path mapping, which easily fits into memory.

Regarding the control plane, the additional network overhead is as low as one hash per on-path AS due to the use of detachable extensions in PCBs.
The inter-domain network overhead for fetching detachable extension data and policy descriptions is greatly reduced through caching at the \emph{local} control service, only requiring intra-AS communication from endpoints.
The remaining overhead, such as the control service configuring intra-AS paths or distributing policies to border routers, depends only on the size of the intra-AS topology and thus scales independently of the global network.

%% file: sections/discussion.tex
\section{Discussion}\label{sec:discussion}
In this section, we discuss limitations and challenges related to the detection of misbehavior and real-world deployment, and potential extension beyond our core design.

\paragraph{Detecting Misbehavior}
We distinguish between two types of router characteristics: (i) quantifiable characteristics that can be measured by endpoints, like hard-/software feature support, and (ii) unquantifiable characteristics that are difficult or impossible to measure, like security properties, location, or jurisdiction.
ASes might not conform to their announced path policies, e.g., due to economic incentives.
Detecting that an AS does not conform to its announced unquantifiable router attributes is very challenging and thus restricts endpoints to rely on, i.e., assume, the correctness of policies announced by trusted ASes as described in \cref{sec:model:trust_model}.
In reality this assumption does not always hold.
For example, some users might not trust their ISPs, in which case \system{}'s objectives are inherently impossible to achieve.
Still, there are many scenarios where the endpoint trusts on-path ASes regarding their unquantifiable router characteristics.
A government agency, for example, is likely to trust providers in the same country, but will not necessarily trust foreign router manufacturers.

For preference policies concerning quantifiable attributes, the trust model can often be relaxed.
In this case, an endpoint can detect misbehavior and switch to a different policy-compliant path.
An endpoint might even be able to pinpoint the misbehaving AS by probing multiple policy-compliant paths using a network tomography approach~\cite{Kakkavas2020}.

\paragraph{Deployment}
A large number of network protocols have never seen major real-world deployment despite their usefulness.
RFC~8170~\cite{RFC8170} contains several design recommendations for the deployment and transition of new network protocols: clear incentives for early adopters, an incremental deployment model, estimation of the total cost including a way to bill benefactors, and extensibility.
\System{} must also overcome these deployment challenges to achieve widespread global adaption.
Both the incremental deployment (\cref{sec:system:policy-dissemination}) and extensibility (\cref{sec:policy_specification}) of \system{} have been discussed already.
ASes should be able to bill the endpoints profiting from policy-enabled paths, for example based on an on-demand or flat-rate model.
One possible solution is for an AS to bill its neighboring AS based on the quantity and quality of policy-compliant paths consumed by the aggregate of all flows.
An endpoint is then billed by its own provider or the provider offers this service for free to attract customers.
A more elaborate approach is to let endpoints directly purchase the use of path policies from the desired ASes, which is more complex and requires prior communication with all on-path ASes, however.
Interestingly, in addition to the mitigations discussed in \cref{sec:security}, billing further helps preventing DDoS attacks as it imposes monetary costs on the attacker.

Since \system{} extends SCION, which is seeing real-world deployment~\cite{anapaya,Kraehenbuehl2021,swisscom}, and it does not require additional capabilities compared to SCION (e.g., SCION border routers already use AES for per-packet operations), existing SCION components do not need to be replaced.
Additionally, the protocols we rely on are already either partially (EPIC~\cite{epic_implemented}) or fully (DRKey~\cite{drkey_implemented}) implemented in SCION\@.
However, an AS deploying \system{} must first become SCION-enabled, which requires changes to its inter-domain routing and its border routers.
Additionally, endpoint devices may lack capabilities, such as native AES support.

Finally, in addition to the overhead discussed in \cref{sec:evaluation}, large ASes with complex internal networks and many internal and border routers may incur additional management overhead for keeping a precise inventory, deciding on which topology information to reveal, and incorporating their traffic engineering policies.

\paragraph{Bidirectional Policy Selection}
In \system{}, an endpoint chooses the policy on the forward path, while the destination endpoint chooses the policy on the backward path.
In some cases, an endpoint may wish to select both a forward and a backward policy-compliant path.

This can for example be achieved through a negotiation protocol, or, if the destination endpoint has no policy preferences (e.g., a publicly available service), by letting the source endpoint dictate the policies to be used by the destination endpoint.
Because the policy indices are encrypted, the endpoints must trust each other as they can not verify the policies chosen by their peer in the headers of the received packets.

One possible mechanism to ensure policy compliance in both directions is for the source endpoint to piggyback the policy indices for the return path in its packets, where the indices are authenticated and encrypted using a shared symmetric key between the end hosts provided by DRKey.
With an additional flag, the source endpoint can signal the destination not to send any return packets in case it does not want to use those indices. This approach requires identical forward and return paths, but guarantees that traffic on the return path (i) traverses source-trusted ASes only and (ii) is compliant with the source's policies.
Working out this mechanism in full detail and evaluating other
alternatives for achieving bidirectional policy compliance remain as future work.

%% file: sections/relatedwork.tex
\section{Related Work}

Platypus~\cite{Raghavan2004} is a source-routing protocol allowing endpoints to compose paths from multiple Internet routes through intermediate waypoints.
Platypus enables fine-grained control over the forwarding path even for intra-AS routes.
This approach is undesirable for ASes that do not want to disclose their internal topology.
Moreover, Platypus requires an endpoint to readily know the desired waypoints.

Alcatraz~\cite{Asoni2018} prevents malicious exfiltration, alteration, and forwarding of data on network devices by leveraging trusted execution environments provided by Intel SGX.
Alcatraz assumes an environment controlled by a single operator.
The throughput achieved on routers was below \SI{1.5}{Gbps} per core, preventing use of Alcatraz in corporate networks and data centers.
Alcatraz could potentially be used as an intra-AS data plane for \system{} according to predefined rules based on the path policies of the AS.

A recent initiative for a responsible Internet~\cite{Hesselman2020} proposes an architecture to increase the transparency, accountability, and controllability of network traffic in the Internet.
The architecture is based on two decentralized systems: the Network Inspection Plane (NIP), which describes how data is handled by different network operators, and the Network Control Plane (NCP), which allows users to influence the treatment of their data in the network.
Although the responsible Internet addresses a different problem, in particular the lack of sovereignty of Europe's network infrastructure, their goal is similar to ours, namely the need for routing over trustworthy network infrastructure.
As mentioned in their work, SCION could be used as a building block to implement parts of the responsible Internet, but has several shortcomings regarding the path selection granularity and global visibility.
Regarding the NCP, \system{} solves some of these shortcomings by enabling more fine-grained path selection related to generic router properties, such as a router's manufacturer, legislation, or software.
Furthermore, \system{}'s AS policies could be consumed by the global NIP and used as a valuable input for policy makers regarding the ASes' intra-domain routing policies and thus enhance transparency in the Internet.

Trusted Path Routing (TPR)~\cite{voit-rats-trustworthy-path-routing-06} allows to enforce that sensitive traffic traversing a network is forwarded only through trustworthy devices.
What comprises sensitive traffic is specified through IP address ranges, which are associated with a trusted topology.
To decide which devices to include in this topology, adjacent devices equipped with a Trusted Platform Module (TPM) mutually verify their trustworthiness through remote attestation~\cite{I-D.ietf-rats-ar4si}.
TPR could complement \system{}, i.e., to provide intra-AS forwarding over a trusted topology consisting of policy-compliant attested routers.\footnote{We verified this by deploying four TPR-enabled CISCO NCS 540 routers~\cite{ncs540} inside a SCION AS. We changed the trustworthiness vector of one of those routers and observed that SCION traffic gets successfully re-routed, thus avoiding the untrustworthy device.}

Besides EPIC, there are several other systems for source authentication and path validation.
OPT~\cite{DRKey} and ICING~\cite{icing} have a lower goodput ratio (ratio between goodput and throughput) compared to EPIC, which is due to longer packet header fields.
Furthermore, ICING causes significantly higher processing overhead at routers than EPIC.
PPV~\cite{ppv} does not provide path validation to the source but enables the destination to probabilistically validate parts of the path.
Similarly, with Hummingbird~\cite{hummingbird}, routers only sample packets probabilistically but based on symmetric keys shared between neighboring routers.
MASK~\cite{mask} only authenticates the packet's source to a single on-path router.

%% file: sections/conclusion.tex
\section{Conclusion}
Properties desired from on-path forwarding devices are inherently subjective:
Some applications require specific router capabilities, while others want to forward traffic only along trusted manufacturer devices, as they might not consider encryption alone sufficient for confidentiality.
This motivates the need for (i) transparent, attested, and non-repudiable claims about on-path device properties, (ii) path selection by individual users, end hosts, applications, or even flows, as well as (iii) protection against hijacking attacks.

The main security takeaways of this work are that the above properties can be achieved by using an expressive policy language on router properties, without compromising AS and endpoint privacy, performance, and scalability.
Namely, we design and implement a system for inter-domain router-based path selection, including a flexible policy language allowing individual router policies per endpoint, that achieves these properties by leveraging the security and extensibility of SCION, remote router attestation, and efficient path validation.
\System{} creates exciting opportunities for ISPs (e.g., creation of new business models and services) as well as for end users (e.g., path selection based on fine-grained policies).

%% file: sections/acknowledgment.tex
\section*{Acknowledgments}
We would like to thank Ralph Mohnhaupt, Bernard Botteron, Chennakesava Reddy Gaddam, Eric Voit, Rakesh Kandula, Vitus Andreoli, Annu Singh, and Georg Aebi from CISCO for providing us with TPR-capable routers and for their support regarding the setup of the measurement testbed.
We further thank Juan Angel García-Pardo for integrating the routers into our network.
We are grateful to the USENIX Security '23 reviewers for their helpful feedback on the manuscript.
We gratefully acknowledge support from ETH Zurich, the Zurich Information Security and Privacy Center (ZISC), armasuisse Science and Technology, and the Werner
Siemens-Stiftung (WSS) Centre for Cyber Trust at ETH Zurich.

%% file: sections/appendix.tex
\section{Questionnaire Results}\label{sec:appendix:questionnaire}
In this section, we present additional insights on our questionnaire, which was approved by the ethics commission of ETH Zürich (2021-N-219).
The questionnaire was distributed to anonymous participants through various channels such as mailing lists.
All answers were collected between April and July 2022 and aggregated to preserve anonymity.
The results are presented in this section and in \cref{sec:survey}.

\begin{figure}[t]
  \includegraphics[width=\linewidth]{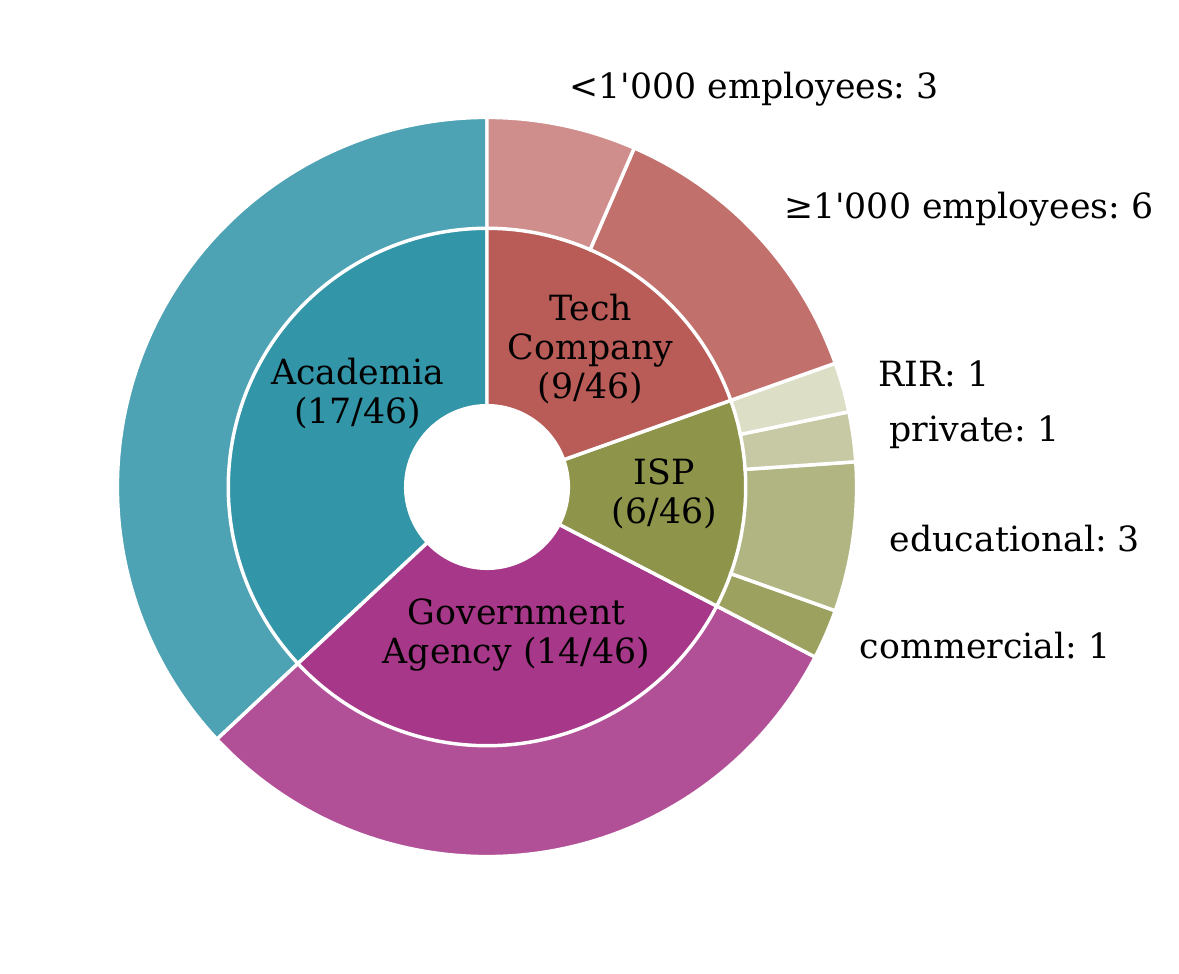}
  \caption{The background of the questionnaire participants.
  }\label{fig:questionnaire:background}
\end{figure}

\Cref{fig:questionnaire:background} shows the distribution of the participants' backgrounds.
We can separate the participants into four groups: tech companies (small, medium, and large), ISPs (commercial, private, and educational ISPs and RIRs), government agencies, and academia (i.e., researchers).

\begin{figure*}[t]
  \captionsetup[subfigure]{justification=raggedleft,singlelinecheck=false,margin=0.6cm}
  \begin{subfigure}[t]{.58\linewidth}
    \includegraphics[height=7cm]{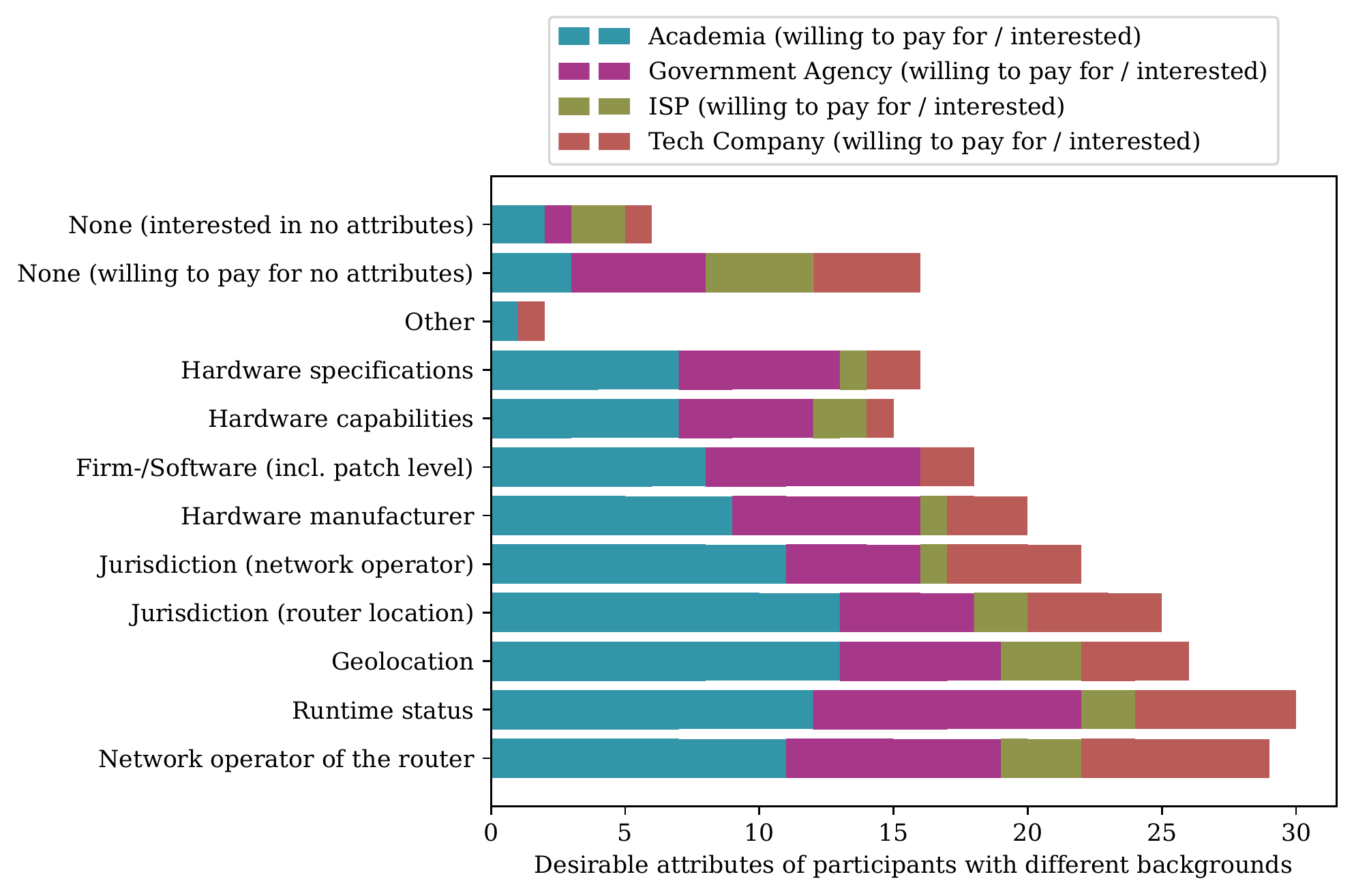}
    \caption{Viewpoint of private customers.}\label{fig:questionnaire:private-background}
  \end{subfigure}
  \hspace{1mm}
  \captionsetup[subfigure]{justification=raggedright,singlelinecheck=false,margin=0.3cm}
  \begin{subfigure}[t]{.40\linewidth}
    \includegraphics[height=7cm]{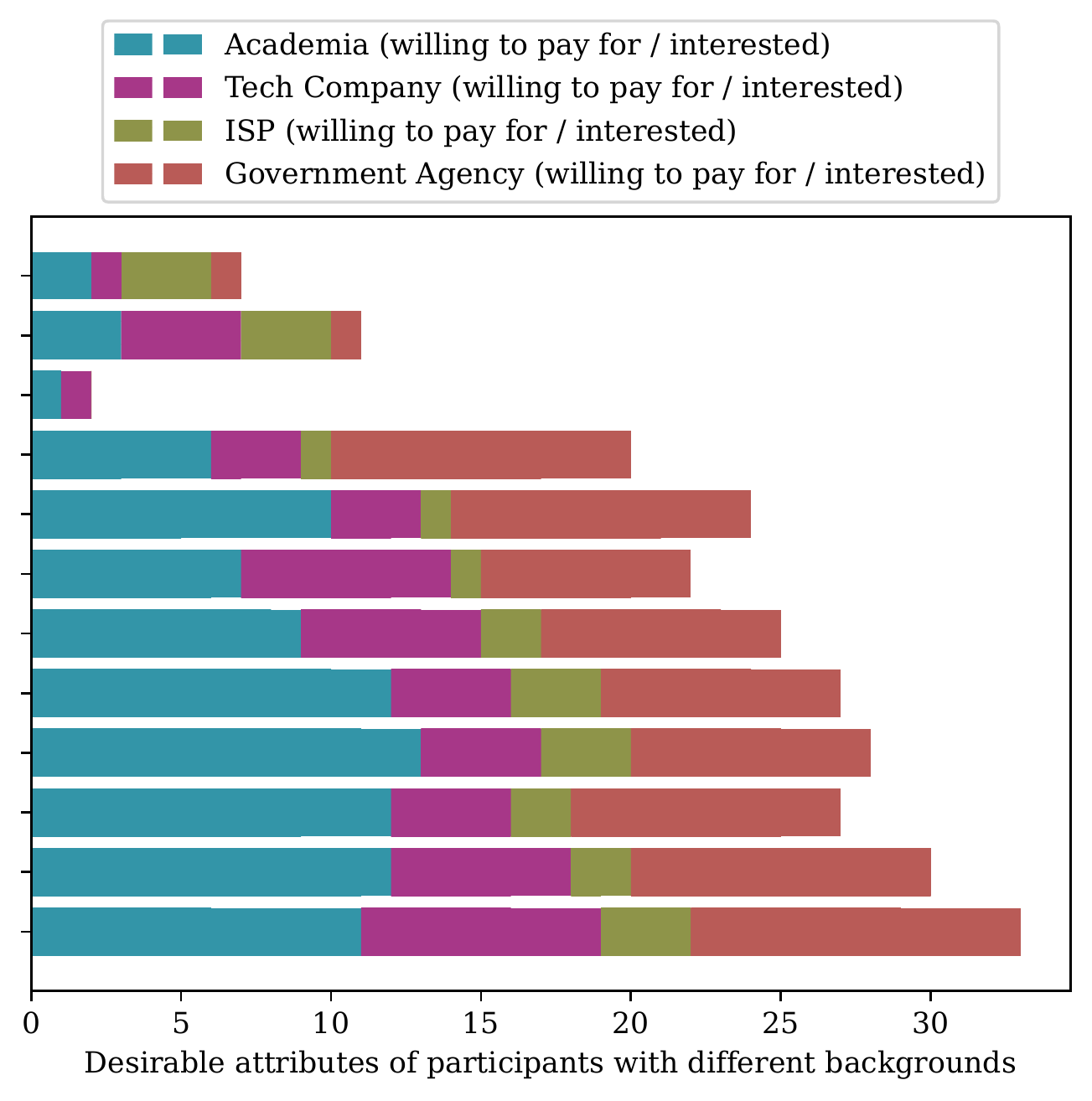}
    \caption{Viewpoint of commercial customers.}\label{fig:questionnaire:commercial-background}
  \end{subfigure}
  \caption{Number of participants with a specific background that are interested in, or willing to pay for, a certain router attribute, respectively.
    \ref{fig:questionnaire:private-background} relates to private customers and \ref{fig:questionnaire:commercial-background} relates to commercial customers.}\label{fig:appendix:questionnaire:per-background-data}
\end{figure*}

\Cref{fig:appendix:questionnaire:per-background-data} shows a detailed breakdown of the data in \cref{fig:questionnaire} indicating which participant groups are interested in which router attributes.
We highlight some relevant findings:
\paragraph{Private vs. Commercial Customers}
Depending on the background, the interest of private customers differs significantly.
In general, the viewpoints of private customers have more variation than that of commercial customers, likely due to personal preferences.
Viewpoints from commercial customers are more aligned among participants with different backgrounds, possibly due to a similar understanding of the requirements of their organization.
\paragraph{Willingness to Pay}
Over \num{90}\% of participants with a government agency background would be willing to pay for influencing the router attributes of on-path routers for commercial customers, which is much higher than participants with tech company and ISP backgrounds (both $<\SI{60}{\percent}$).
This might be caused by the more stringent security and compliance requirements of government agencies.
\paragraph{Inconsistencies} Surprisingly, four participants claimed to be willing to pay for information about certain router attributes while not mentioning that they are interested in these attributes.
This might be due to a willingness to pay not because of interest but rather for compliance reasons, or simply due to inattention when filling out the survey.

\section{DRKey}\label{sec:background:drkey}
To encrypt and authenticate endpoint-selected policies, we rely on the fast key derivation provided by DRKey, which in turn enables us to leverage and extend the per-packet cryptographic operations introduced in EPIC.

The DRKey~\cite{DRKey, PISKES} system allows ASes and endpoints to efficiently exchange symmetric keys. To make the symmetric keys issued by AS~$A$ available to its infrastructure components, such as border routers, DRKey does not store the keys at those components, but instead enables the components to recompute the keys efficiently on the fly.
For this purpose, AS~$A$ keeps a single secret (symmetric) key \key{A}{} known by its relevant components (the control service and all border routers). When it receives a key request by AS~$B$, it responds with an AS-level symmetric key \key{A \rightarrow{} B}{} computed with a pseudorandom function (PRF):
\begin{equation}
    \key{A \rightarrow{} B}{} := \prf{\key{A}{}}{B}.
\end{equation}
Upon receiving this AS-level key, AS~$B$ can use it to further derive symmetric keys for its endpoints. It computes symmetric keys between endpoint~$H_B$ (in AS~$B$) and AS~$A$, and between endpoint~$H_B$ and endpoint~$H_A$ (in AS~$A$) as follows:
\begin{align}
    \key{A \rightarrow{} B:H_B}{} &:= \prf{\key{A \rightarrow{} B}{}}{H_B},\label{eq_key_as_host}\\
    \key{A:H_A \rightarrow{} B:H_B}{} &:= \prf{\key{A \rightarrow{} B}{}}{H_A, H_B}.
\end{align}
Here, the comma separating the endpoint identifiers denotes concatenation, and the colons in the key names make explicit to which AS an endpoint belongs to.
For traffic going through a border router of AS~$A$ and originating from endpoint~$H_B$, the router can recompute \key{A \rightarrow{} B:H_B}{} based solely on its secret key \key{A}{}, plus the AS-identifier~$B$ and the endpoint source address~$H_B$, which are part of the SCION packet header.
For communication with destination endpoint~$H_A$, the source endpoint~$H_B$ has to explicitly request \key{A:H_A \rightarrow{} B:H_B}{} from the control service of AS~$B$, as it does not have the necessary AS-level key \key{A \rightarrow{} B}{} to derive it autonomously.
Similarly, if the endpoint~$H_A$ wants to know \key{A:H_A \rightarrow{} B:H_B}{} for packets received from $H_B$, it must request this key from its local control service of AS~$A$, as $H_A$ is neither in possession of \key{A}{} nor \key{A \rightarrow{} B}{}.
Endpoint and AS identifiers are not secret, but to compute \key{A:H_A \rightarrow{} B:H_B}{}, the key \key{A \rightarrow{} B}{} is needed, which can either be derived directly at border routers and the control service of AS~$A$ from \key{A}{}, or it needs to be fetched by AS~$B$ from the control service of AS~$A$.

\section{EPIC}\label{sec:background:epic}
Leveraging DRKey's efficient key-derivation, EPIC~\cite{legner2020epic} enables per-packet source authentication and path validation in SCION. 
\emph{Source authentication} at border routers protects both the network and the destination endpoint, as it allows filtering unauthentic packets early and before they reach any bottleneck links.
Through \emph{path validation}, an endpoint can verify whether its traffic indeed followed the selected path on the AS-level.

To authenticate a source endpoint~$H_S$ in AS~$A_0$ to on-path border routers, the source endpoint computes a per-packet hop validation field (HVF) for each AS~$A_i$ ($1 \leq i \leq n$) on the selected path, which it includes in the SCION packet header:
\begin{align}
    \key{i}{} &:= \key{A_i \rightarrow{} A_0:H_S}{} \label{eq_epic_key_abbr} \\
    \hvf{i} &:= \mac{\key{i}{}}{\tspkt{}, A_0, H_S, \sigma_i}~\text{[0:\lval]} \label{eq_epic_hvf}
\end{align}
Here, the function MAC$_K$ computes a message authentication code with key $K$ and \tspkt{} denotes a high-precision timestamp added to the packet header that is unique for every packet sent by source endpoint~$H_S$.\footnote{The last input to the MAC, $\sigma_i$, is an authenticator included by EPIC to achieve a property called \emph{path authorization}, which protects the routing decision of ASes from malicious endpoints. The computation of $\sigma_i$ is irrelevant for our work.}
The notation X[a:b] denotes the substring from byte a (incl.) to byte b (excl.) of X, and the HVF is thus defined as the first \lval{} bytes of the MAC output.
To verify a packet's source, a border router recomputes the HVF of its AS and compares it to the one contained in the packet header.

To provide path validation to the source endpoint, the border routers replace the HVFs in the packet with the next \lval{} bytes of the MAC output, i.e., [\lval:2\lval]. This serves as proof that the packet indeed traversed the ASes on the selected path. To communicate this information to the source endpoint, the destination endpoint~$H_D$ returns a packet containing the updated HVFs and \tspkt{} of the original packet, authenticated with its symmetric key \key{A_\ell:H_D \rightarrow{} A_0:H_S}{}.

Through the deployment of a duplicate-suppression system, EPIC furthermore allows ASes to filter replayed packets. Also, an EPIC-enabled source endpoint authenticates the whole packet including the payload for the destination endpoint, and therefore also includes a corresponding destination validation field in the packet header.

\section{Securing Path Policy Indices}\label{sec:appendix:computation}
To provide authenticity and confidentiality of the policies selected by endpoints, we encrypt the corresponding indices and add them to the input of the EPIC per-packet authenticator computation from \cref{eq_epic_hvf}:
\begin{align}
    \ind{i}{E} &= \ind{i}{} ~\oplus~ (\aes{\key{i}{}}{\tspkt{}}~\text{[0:\lpol]}) \label{eq_enc_new} \\
    \hvf{i} &= \mac{\key{i}{}}{\tspkt{}, A_0, H_S, \sigma_i, \ind{i}{E}}~\text{[0:\lval]} \label{eq_hvf_new}
\end{align}
Here, \ind{i}{} denotes the plaintext and \ind{i}{E} the encrypted policy index for on-path AS~$i$, respectively.
The length of the policy index in bytes is given by \lpol{}. 
Again, the notation X[a:b] denotes the substring from byte a (incl.) to byte b (excl.) of X, and the comma-separated inputs to the MAC function are concatenated.
This approach follows the Encrypt-then-MAC idea, where we use the CTR mode for encryption and where the timestamp \tspkt{} serves as nonce. The counter is always zero, i.e., omitted, because the length of the input (\tspkt{}) is shorter than the AES block size.
Alternatively, the computation of \ind{i}{E} in \cref{eq_enc_new} can be understood as one-time pad encryption using the key \aes{\key{i}{}}{\tspkt{}}.
The source endpoint adds \tspkt{}, \ind{i}{E} and \hvf{i} for every on-path AS~$i$ to its data packet.

The reason we do not directly apply the AES block cipher to the packet timestamp and the policy index (i.e., without using the bitwise XOR operation) is because this would result in an encrypted policy index with a size of \SI{16}{\byte} (AES block size).
With our approach, the encrypted policy index consists of only \lpol{} bytes, which allows for shorter packet headers.

Upon reception of a data packet, the ingress border router of the $i$-th on-path AS checks that the timestamp is current, derives \key{i}{} (Equations \ref{eq_key_as_host} and \ref{eq_epic_key_abbr}), and re-computes \hvf{i} (Equation~\ref{eq_hvf_new}) and compares it to the \hvf{i} contained in the packet. If they do not match, the packet is dropped. Otherwise, the router decrypts \ind{i}{Enc} using $\oplus$ to obtain the policy index \ind{i}{}. It then replaces the HVF with the the \lval{} next bytes of the MAC output and, based on a local lookup table, forwards the packet such that it traverses the intra-AS network in a policy-compliant manner. In case there is no entry in the table for this policy index, the router sends back a control message to the source endpoint and drops the packet.
The destination endpoint is not modified, it checks the destination validation field and returns an authenticated confirmation containing the updated HVFs and \tspkt{} of the original packet, such that the source can verify the AS-level path the packet traversed.

Through those modifications, we extend EPIC to not only achieve source authentication and path validation, but also secrecy and authenticity of the policy indices.
Because policy indices are encrypted on a per-packet basis, an attacker cannot infer whether two encrypted indices describe the same plaintext policy index by only looking at the packet headers.
Hence, such an attacker can not even deduce whether the source endpoint has chosen any policies at all, as a policy index of zero is indistinguishable from any other policy index after encryption.

\section{Policy Specification Details}
In this section, we provide additional details on the syntax of the policy language and our policy encoding.

\subsection{Policy Syntax}\label{sec:appendix:policy-syntax}
The syntax of our policy language is based on the following general syntax for FOL with equality. In addition to the sorts, function symbols, and predicates defined in \cref{sec:policy-syntax}, we have

\balance

\begin{itemize}[label={}]
\item \textbf{Sorts}: A set of sorts $\{I_1, I_2, ..., I_n\}$
\item \textbf{Vars}: A set of sorted variables $\{v_1, v_2, ...\}$, where each variable $v_k$ is of some sort $I$, written $v_k^I$.
\item \textbf{Function Symbols}: A set of function symbols.
  The input parameters and the output of a function (i.e., its signature) are of specific sorts: $f: I_{i_1} \times I_{i_2} \times ... \times I_{i_n} \rightarrow I_r$
\item \textbf{Constants}: A set of sorted constants (i.e., function symbols with arity 0)
\item \textbf{Predicates}: A set of predicates.
  The input parameters of an $n$-ary predicate are of the sorts $I_{i_1}, I_{i_2}, ..., I_{i_n}$, where $n\geq 1$
\item \textbf{Terms} are defined inductively:
  \begin{itemize}[label={}]
  \item Vars $\subseteq$ Terms
  \item Constants $\subseteq$ Terms
  \item if $t_1^{I_{t_{1}}}, t_2^{I_{t_{2}}}, ..., t_k^{I_{t_{k}}} \in \text{Terms}$ and $f$ is a function with signature $I_{t_1} \times I_{t_2} \times ... \times I_{t_k} \rightarrow I_{r}$, then $f(t_1, t_2, ..., t_k)^{I_{r}} \in \text{Terms}$
  \end{itemize}
\item \textbf{Formulas} are defined inductively:
  \begin{itemize}[label={}]
  \item if $t_1^I, t_2^I \in \text{Terms}$ for some sort $I$, then $t_1 = t_2 \in \text{Formulas}$
  \item if $t_1^{I_1}, t_2^{I_2}, ..., t_k^{I_k} \in \text{Terms}$ and $P$ is a predicate with signature $I_1 \times I_2 \times ... \times I_k$, then $P(t_1, t_2, ..., t_k) \in \text{Formulas}$
  \end{itemize}
\item if $\phi \in \text{Formulas}$ and $\theta \in \text{Formulas}$, then
  \begin{itemize}[label={}]
  \item $\neg \phi,\enspace \phi \wedge \theta,\enspace \phi \vee \theta,\enspace \phi \rightarrow \theta \enspace \in \text{Formulas}$
  \end{itemize}
\item if $\phi \in \text{Formulas}$ and $x^I \in \text{Vars}$ is a variable of sort $I$, then
  \begin{itemize}[label={}]
  \item $\forall x^I: \phi,\enspace \exists x^I: \phi \enspace \in \text{Formulas}$
  \end{itemize}
\item if $t_1^I, t_2^I \in \text{Terms}$ are terms of sort $I$ (note that in our policy language, inequality is only defined for sort $V$), then
  \begin{itemize}[label={}]
  \item $t_1 < t_2,\enspace t_1 \leq t_2,\enspace t_1 > t_2,\enspace t_1 \geq t_2 \enspace \in \text{Formulas}$
  \end{itemize}
\end{itemize}

\subsection{Semantics and Interpretation}\label{sec:appendix:semantics}
An interpretation for our sorted FOL with equality is defined by a set of carrier sets interpreting each sort, a set of relations interpreting each predicate, and a set of functions (including constants with arity 0) interpreting each function symbol.
We first define interpretation of the sorts in our language:
\begin{itemize}[leftmargin=0pt,label=]
    \setlength\itemsep{0pt}
    \item \textbf{Manufacturer (Sort $\boldsymbol M$):} Entity creating and distributing routers. Let $\mathcal{M}$ be the set of all manufacturers.
    \item \textbf{Software Component (Sort $\boldsymbol C$):} A software product, which is identified by a unique tag, typically assigned by the owner of the software product, and may support versioning.
    Let $\mathcal{C}$ be the set of all software components.
    \item \textbf{Tag (Sort $\boldsymbol T$):} A software component is clearly identified by a globally unique tag.
    Let $\mathcal{T}$ be the set of all possible tags.
    The function \texttt{tag}: $\mathcal{C} \rightarrow \mathcal{T}$ returns the tag of a software component.
    \item \textbf{Tag Issuer (Sort $\boldsymbol I$):} The tag of a software component is assigned by a tag issuer, which is identified by a URI.
    Let $\mathfrak{I}$ be the set of all possible issuers.
    The function \texttt{issuer}: $\mathcal{T} \rightarrow \mathfrak{I}$ returns the issuer of a tag.
    \item \textbf{Name (Sort $\boldsymbol N$):} The name of a software component, which allows the identification of a software component with different versions.
    Let $\mathcal{N}$ be the set of all possible names assigned to a software component.
    The function \texttt{name}: $\mathcal{C} \rightarrow \mathcal{N}$ returns the name of a software component.
    \item \textbf{Version (Sort $\boldsymbol V$):} A software component may be identified by a specific version.
    Let $\mathcal{V}$ be a totally ordered set of all possible version numbers for different version schemes defined in the NIST standard for Software Identification Tags~\cite[5.1.2]{nist-8060-swid}.
    The function \texttt{version}: $\mathcal{C} \rightarrow \mathcal{V}$ returns the version of a software component or the default version (``1.0.0'').
    \item \textbf{Router (Sort $\boldsymbol R$):} A device produced by a specific manufacturer and running a clearly defined software stack.
    Let $\mathcal{R}$ be the set of all possible routers.
    \item \textbf{Path (Sort $\boldsymbol P$):} A sequence of routers but we simplify the representation of a path as a set since order and repetition are not relevant for path policies.
    The elements of a path are thus defined by the relation \texttt{onPath}: $\mathcal{P} \times \mathcal{R}$.
    Let $\mathcal{P}$ be the set of all possible paths.
    \item \textbf{Router Setup:} A router setup consists of its manufacturer and software stack, i.e., the set of software components providing the router functionality.
    It is thus defined by the function and relation, \texttt{manufacturer}: $\mathcal{R} \rightarrow \mathcal{M}$ and \texttt{software}: $\mathcal{R} \times \mathcal{C}$.
\end{itemize}
Based on the above, we define an interpretation as follows:
\begin{itemize}[leftmargin=0pt,label=]
    \setlength\itemsep{0pt}
    \item \textbf{Sorts:} The sorts and their respective carrier sets are defined in \cref{tab:sorts}.
    The concrete values used to encode these router attributes are described in \cref{sec:appendix:policy-encoding}.
    \item \textbf{Predicates:} Predicate symbols are interpreted by relations. The predicate for the total order on $\mathcal{V}$ overloads the comparison operator (relation) $\leq$ over $V \times V$.
    This total order is defined by the version scheme (e.g., SemVar where versions use a three-part version number such that $1.2.3 \leq 1.3.2$).
    The remaining operators $<, \geq$, and $>$ can be defined in terms of $\leq$\footnote{$p < q := p \leq q \wedge p \neq q$, $p \geq q := q \leq p$, and $p > q := q < p$}.
    Finally, there are the relations \texttt{onPath} and \texttt{software}.
    \item \textbf{Functions:} The function symbols \texttt{manufacturer}, \texttt{tag}, \texttt{issuer}, \texttt{name}, and \texttt{version} are interpreted by their respective functions.
    Constants, i.e., function symbols with arity 0, are interpreted by the carrier set of their respective sort as described in \cref{tab:sorts}.
\end{itemize}

\subsection{Policy Encoding}\label{sec:appendix:policy-encoding}
Our policies are based on two building blocks: manufacturers and software components with their respective attributes (tags, issuers, names, and versions).
To communicate policies among different entities, we need a common set of values for each building block.
We encode the manufacturer using the private enterprise numbers specified by IANA~\cite{iana-enterprise-numbers}.
These numbers are globally consistent unique identifiers for different manufacturers.
The software components including their related attributes are encoded using \texttt{SoftwareIdentity} elements from the NIST standard for Software Identification Tags~\cite{nist-8060-swid}.